\documentclass[times, review, 10pt]{elsarticle}

\usepackage{amsmath}
\usepackage{graphicx}
\usepackage{multirow}
\usepackage[mathscr]{euscript}
\usepackage{hyperref}

\DeclareMathOperator{\tr}{Tr}
\newcommand {\vc} [1] {\boldsymbol{#1}}
\usepackage{amsfonts,amsthm,bm, lipsum}
\usepackage{adjustbox}
\usepackage{textcomp}

\usepackage{subcaption}

\begin{document}

\begin{frontmatter}

\makeatletter
\def\theaffn{\arabic{affn}}  
\makeatother

\title{SUPER-Net: Trustworthy Image Segmentation via Uncertainty Propagation in Encoder-Decoder Networks}

\tnotetext[note1]{This is the author’s accepted manuscript. 
Licensed under CC BY-NC-ND. 
The Version of Record is available at \url{https://doi.org/10.1016/j.patcog.2025.112503}. 
Supplementary material is included as an ancillary file in this arXiv submission.}

 \author[1]{Giuseppina Carannante}
 \author[1]{Nidhal~C. Bouaynaya}
 \author[2]{Dimah Dera}
  \author[3]{Hassan~M. Fathallah-Shaykh}
 \author[4]{Ghulam Rasool}
 \affiliation[1]{organization={Department
of Electrical and Computer Engineering, Rowan University},
             city={Glassboro},
            state={NJ},
            country={USA}}
 \affiliation[2]{organization={Chester F. Carlson Center for Imaging Science, Rochester Institute of Technology},
             city={Rochester},
            state={NY},
            country={USA}}

  \affiliation[3]{organization={Department of Neurology, University of Alabama at Birmingham School of Medicine},
             city={Birmingham},
            state={AL},
            country={USA}}

  \affiliation[4]{organization={Machine Learning Department, Moffit Cancer Center},
             city={Tampa},
            state={FL},
            country={USA}}

\begin{abstract}
Deep Learning (DL) holds great promise in reshaping the industry owing to its precision, efficiency, and objectivity. However, the brittleness of DL models to noisy and out-of-distribution inputs is ailing their deployment in sensitive fields. Current models often lack uncertainty quantification, providing only point estimates. We propose SUPER-Net, a Bayesian framework for trustworthy image segmentation via uncertainty propagation. Using Taylor series approximations, SUPER-Net propagates the mean and covariance of the model's posterior distribution across nonlinear layers. It generates two outputs simultaneously: the segmented image and a pixel-wise uncertainty map, eliminating the need for expensive Monte Carlo sampling. SUPER-Net’s performance is extensively evaluated on MRI and CT scans under various noisy and adversarial conditions. Results show that SUPER-Net outperforms state-of-the-art models in robustness and accuracy. The uncertainty map identifies low-confidence areas affected by noise or attacks, allowing the model to self-assess segmentation reliability, particularly when errors arise from noise or adversarial examples.
\end{abstract}



\end{frontmatter}

\section{Introduction}
\label{sec1}
Driven by the superior performance achieved in many areas, various deep learning (DL) models have been advanced to analyze medical data, e.g., radiological images and pathology slides.
Several methods have achieved, if not surpassed, prognosis parity with specialized medical personnel \citep{liu2019comparison}. However, their successful deployment in clinical settings remains limited. While several autonomous algorithms are doubtlessly employed for many everyday tasks --- e.g., spam filters for emails or biometrics that unlock our cellphones ---, there is a less assertive willingness to utilize the same algorithms for risky, sensitive data, such as medical images.

The main challenge that hinders the widespread and effective use of DL in clinical settings is the lack of reliable and trustworthy predictions \citep{biggio2018wild}. For example, when encountering test examples that differ significantly from its training data, a DL system will still produce a prediction. However, without uncertainty information, there is no way to determine how reliable that prediction is. 
This concern is further exacerbated by the vulnerability of DL models to adversarial inputs --- perturbations that are imperceptible to human observers yet cause a trained DL model to produce erroneous predictions \citep{goodfellow2015explaining}. In the literature, there are studies highlighting the vulnerability of medical models to adversarial perturbations  \citep{finlayson2019adversarial}. 
As a result, DL in medicine is particularly susceptible due to both technical weaknesses and financial incentives \citep{finlayson2019adversarial}. 

Addressing these challenges requires DL models not only to produce accurate predictions but also to quantify the uncertainty associated with those predictions. Uncertainty Quantification (UQ) serves as a key mechanism for assessing the reliability of predictions, allowing users to be aware of the level of confidence in the models' predictions. UQ could be very useful when the DL model is essentially guessing at random due to excessive noise in the input or possible adversarial attacks.
Unfortunately, as most DL models are inherently deterministic, a measure of confidence or uncertainty is not readily available at their output. 

Estimating the confidence of a model requires a probabilistic interpretation of the model's parameters, i.e., treating model parameters as random variables endowed with a probability distribution. Through Bayesian inference, the posterior distribution of the model parameters can be found. At test time, the second moment, i.e., the covariance, of the predictive distribution can serve as a measure of confidence or uncertainty in the predicted output.
Several Bayesian models have been developed for the classification and regression problems \citep{goan2020bayesian}. Trade-offs between prediction accuracy, confidence estimation, and scalability are at the heart of these different approaches \citep{goan2020bayesian}. 

A relatively small amount of work focuses on quantifying uncertainty in pixel-level segmentation tasks using Bayesian DL models. The challenge in learning uncertainty for each pixel arises from propagating high-dimensional posterior distributions of the model's parameters through multiple stages of non-linearities in the encoder-decoder architecture. Furthermore, the model must provide an \emph{instantaneous} uncertainty map at test time, i.e., simultaneously output the prediction (the segmentation) and corresponding pixel-level uncertainty map \emph{without} resorting to expensive Monte Carlo sampling techniques or model averaging (ensemble).

Recently, Dera \emph{et al.} proposed a variational moments’ propagation (VMP) framework that provides a meaningful and scalable framework for uncertainty propagation and estimation in Convolutional neural network (CNN) classifiers \citep{dera2019extended}.  The mathematical derivations presented in \citep{dera2019extended} are not sufficient for uncertainty propagation in encode-decoder-based segmentation neural networks. The challenges in adopting the VMP framework in segmentation lie in the nature of the learning task: semantic segmentation requires the extraction of both global and local contextual information by encoding and then decoding the input data. Consequently, segmentation networks are fundamentally different than classification networks, e.g., CNNs. The mathematical derivations for the decoder part were never presented previously, and the flow of uncertainty from the encoder to the decoder was never considered within an analytical and systematic framework. The decoder presents specific non-linearities and operations, e.g., up-sampling, padding, and concatenation, that require new mathematical derivations to track the propagation of the uncertainty.  
In addition, our previous work estimated a scalar (or a vector) value of variance that is associated with the predicted class \citep{dera2019extended}. This work introduces the notion of a dense, pixel-level uncertainty map that is provided simultaneously along with the predicted segmentation.

In this paper, we develop a VMP framework for segmentation tasks, SUPER-Net, and extensively evaluate it for various medical imaging datasets under various noisy and adversarial conditions. By leveraging key concepts from probability density tracking in nonlinear and non-Gaussian systems \citep{doucet2009tutorial}, we propagate the first and the second moments of the posterior distribution of network parameters through the nonlinear layers of an encoder-decoder type segmentation architecture. The developed approach is tested using various medical segmentation datasets consisting of Magnetic Resonance Images (MRIs) and Computed Tomography (CT) scans. The proposed VMP formulation and the derived mathematical relationships presented in the paper are applicable to various DNN architectures.

The contributions of this paper are summarized as follows: 

(1) Formalize a scalable Bayesian framework that \emph{simultaneously} learns pixel-wise prediction and confidence in encoder-decoder segmentation networks by analytically approximating and maximizing the evidence lower bound (ELBO). Using first-order Taylor series approximation, we derived closed-form expressions to propagate the first two moments (mean and covariance) of the posterior distribution of the model parameters given the training data and update them during backpropagation; thus, effectively learning the intrinsic uncertainty of the model. We derive mathematical relations for all operations involved, rendering a method that is adaptable to other models, e.g., Variational Autoencoders, and to other tasks as well.

(2) Develop a Bayesian DL architecture that instantaneously outputs two maps: (1) the segmented image and (2) the uncertainty map of the predicted segmentation. These two maps are delivered simultaneously and \emph{without} requiring any Monte Carlo sampling at inference time. That is, the generated uncertainty was intrinsically learned by the model rather than estimated post-training.

(3) Extensively evaluate the performance of the proposed SUPER-Net for various medical segmentation tasks and under various signal-to-noise ratios (SNRs) and conditions. A thorough robustness analysis is conducted by assessing the performance of the model and uncertainty map under these perturbations of the input data.

\section{Related Work}\label{related}
Image segmentation is a fundamental problem in computer vision with applications ranging from medical image analysis to scene understanding for autonomous vehicles. DL techniques, particularly Fully Convolutional Networks (FCNs), have been widely used for pixel-level segmentation \citep{long2015fully}. FCNs modify traditional CNN architectures by replacing fully connected layers with upsampling operations to generate segmentation masks.
  \emph{Encoder-decoder} architectures have since become the dominant paradigm for semantic segmentation \citep{ronneberger2015u}. The \emph{encoder} extracts low-dimensional (salient) features of the data, while the \emph{decoder} reconstructs the spacial information to perform pixel-wise classification. 
Various improvements have been introduced, e.g., skip connections with attention mechanisms \citep{cao2024rasnet}, dilated convolutions \citep{cahall2019inception}, wide contest blocks or compression extraction modules \citep{rehman2020bu}. 

More recently, Transformer-based architectures have been explored for segmentation, leveraging attention mechanisms to capture long-range dependencies \citep{yan20233d}.  
Readers interested in further details are directed to recent surveys on the application of Transformers to various segmentation tasks \citep{thisanke2023semantic}, particularly within the medical domain \citep{xiao2023transformers}. Inspired by the success of foundational models in natural language processing, the Segment Anything Model (SAM) \citep{kirillov2023segment} introduced a zero-shot approach to segmentation, which has also been evaluated for medical imaging tasks \citep{huang2024segment}.

These architectural advances, however, focused on improving accuracy which, no doubt, is an important metric but it does not convey the full picture. Reliability, robustness, and trustworthiness are important metrics for these models. An unreliable model can jeopardize the clinical system by exposing it to technical vulnerabilities, financial risks, and even patient harm \citep{finlayson2019adversarial}. 
In the context of semantic segmentation, there are two main approaches for UQ: Monte Carlo (MC) dropout \citep{gal2016dropout} and model ensemble \citep{lakshminarayanan2017simple}.

MC dropout is widely used due to its simplicity and compatibility with existing NN architectures \citep{gal2016dropout}. The uncertainty information is obtained, at inference time, from the sample variance of multiple MC forward passes through the network. Several studies have applied this technique for various segmentation tasks \citep{kendall2017bayesian}. For instance, a full-resolution residual network is used for brain segmentation in \citep{jungo2017towards}, the QuickNAt architecture is used in  \citep{roy2019bayesian}.

In ensemble methods \citep{lakshminarayanan2017simple}, after training multiple networks, usually with random initialization, several segmentation estimates are produced, and their variation is used as a measure of confidence. For example, \citep{larrazabal2021orthogonal} uses a ResUNet architecture with soft dice loss and two regularization terms to diversify the ensemble members. Authors in \citep{kamnitsas2017ensembles} generate diversity in the ensemble by considering predictions generated by different architectures and models. 
Some works combined the two approaches; for example, in \citep{ghoshal2021estimating} ensemble members are generated by changing the dropout rate. 
Other techniques, e.g., hierarchical probabilistic models \citep{baumgartner2019phiseg}, Evidential Deep Learning \citep{li2023region} and Normalized Softmax Entropy  \citep{guo2024uctnet}, have also been explored to quantify uncertainty. However, most existing UQ approaches share common limitations.

Post-hoc methods estimate uncertainty only at inference time—using multiple forward passes or MC sampling—rather than integrating it into training. This prevents the model from refining uncertainty estimates based on training data. Moreover, these methods approximate uncertainty through empirical sample variance, which may not reflect true confidence, leading to overconfident incorrect predictions. They are also computationally expensive, requiring multiple forward passes at test time or training multiple models in ensemble methods. Lastly, many approaches lack robustness evaluation, as they are often assessed on clean datasets without considering adversarial attacks or noisy inputs
\begin{figure*} [t]
\begin{center}
\hspace{0mm}\includegraphics[width=1\textwidth]{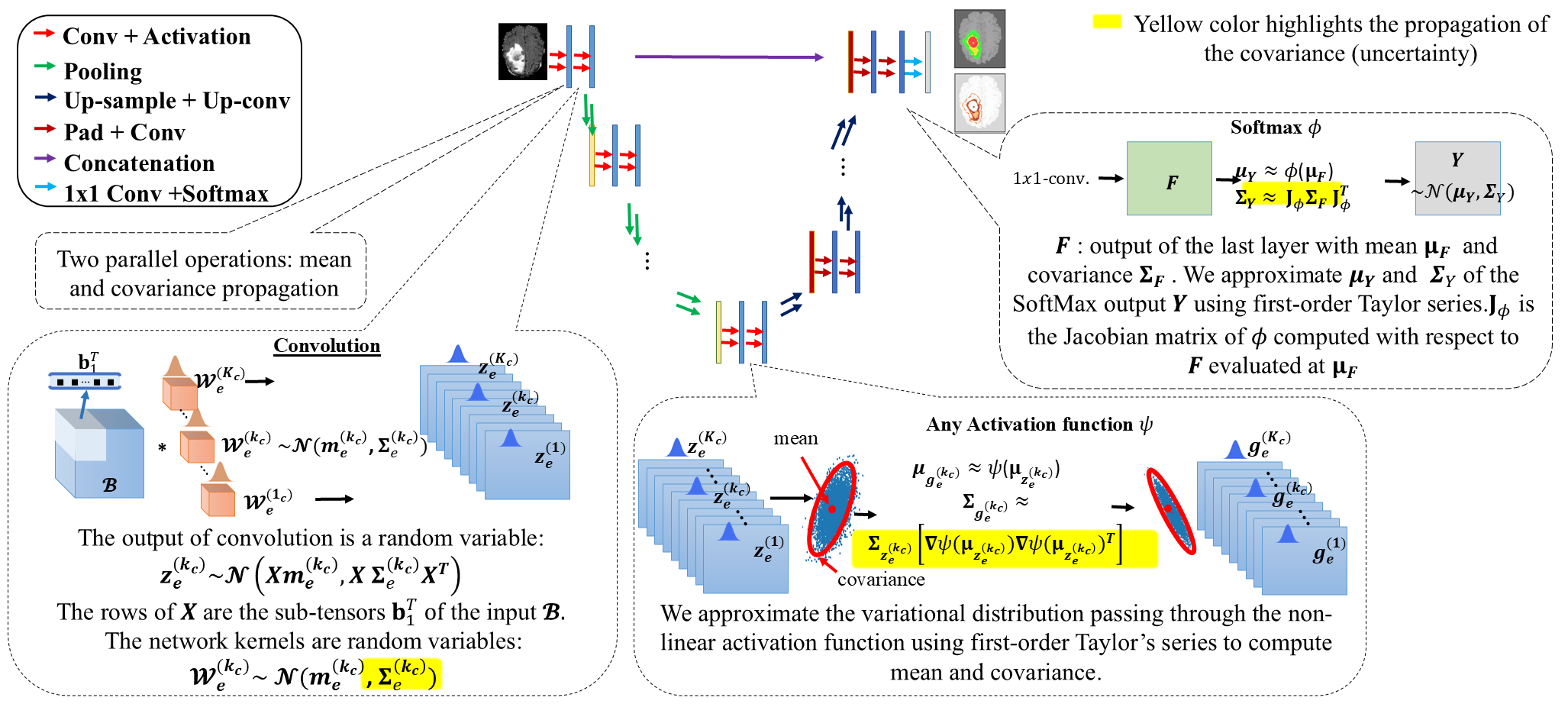}
\caption{An illustration of the SUPER-Net model, where all mathematical operations are performed on random variables. At each step, both the mean and covariance are propagated (indicated by the two arrows). The output of SUPER-Net consists of the predicted segmented image and a covariance matrix, which is used to generate the associated uncertainty map.}\label{fig:fig1}
\end{center}
\end{figure*}

In contrast, the proposed SUPER-Net framework learns uncertainty during training and outputs simultaneously the predicted segmentation and its uncertainty map. A framework to learn the variance is proposed in \citep{dera2019extended}, but derivations are limited to CNNs, rendering the approach unsuitable for the more complex end-to-end segmentation tasks. Building upon this work, we develop a Bayesian framework that propagates the first and the second moment of the variational posterior distribution across all layers of a segmentation DL model. At test time, the uncertainty in the predicted segmentation is produced by the network as the covariance matrix of the predictive distribution simultaneously alongside the segmentation without resorting to multiple runs.

\section{SUPER-Net: Segmentation with Uncertainty Propagation in Encoder-decodeR Networks}\label{method}
We derive a Bayesian framework where the first two moments of the posterior distribution are learned simultaneously by (forward and backward) propagation through the network layers. For scalability and efficiency of the proposed approach, we adopt the Variational Inference (VI) technique, but rather than estimating the expected log-likelihood using expensive Monte Carlo sampling, we approximate its first two moments with a first-order Taylor-series expansion. In the sequel, we present our mathematical results. An illustration of the SUPER-Net model is presented in Figure \ref{fig:fig1}.
\subsection{Mathematical Notations}
Scalars are represented by lower-case letters, e.g., $x$, $x_i$. Vectors are represented by bold lower-case letters, e.g., $\mathbf{y}$. All vectors are column vectors. ${y}_i$ denotes the $i^{\text{th}}$ element of vector $\mathbf{y}$. Matrices are represented by bold upper-case letters, e.g., $\mathbf{A}$. Tr$(\cdot)$ denotes the trace of a matrix, i.e., the sum of its diagonal elements. $^T$ denotes the transpose operator, and vec$(\cdot)$ denotes the vectorization operator. The Hadamard product, i.e., the element-wise product, is denoted with $\odot$, while $\times$ represents the matrix-matrix or matrix-vector product.
Tensors with three or more dimensions are represented by curly bold upper-case letters, e.g., $\bm{\mathcal{X}}$.
If $x$ is a random variable, $\mathbb{E}[x]$ denotes the expected value of $x$. We use $\bm{\mathcal{W}}_{e}^{(k_c)}$ to represent the $k_c^{\text{th}}$ convolutional kernel of the $c^{\text{th}}$ layer. $K_c$ denotes the total number of kernels in layer $c$. The subscripts $e$ and $d$ represent the encoder and decoder path operations, respectively. 
\subsection{Bayesian Deep Learning and Variational Inference}
In Bayesian statistics, the unknown parameters are fully characterized by their posterior distribution given the observations. In Bayesian DL, the network parameters $\mathrm{\Omega}$ are endowed with a prior probability distribution $p(\mathrm{\Omega})$ and all information about the parameters is embedded in the posterior distribution $p(\mathrm{\Omega}|{\cal{D}})$ given the (training) data $\mathcal{D}= \{\bm{\mathcal{X}}^{i},\mathbf{y}^{i}\}_{i = 1}^{N}$. Once the posterior is estimated, the predictive distribution, i.e., the distribution of the test data, can be derived as:
\begin{equation} \label{eq:1}
p(\mathbf{y}^{*}|\bm{\mathcal{X}^{*}},\mathcal{D})= \int p(\mathbf{y}^{*}|\bm{\mathcal{X}^{*}},\Omega)~ p(\Omega|\mathcal{D})~ d\Omega,
\end{equation}
\noindent where $\bm{\mathcal{X}^{*}}$ is the input, $\mathbf{y}^{*} $ is its corresponding predicted output and $p(\mathbf{y}^{*}|\bm{\mathcal{X}^{*}},\Omega)$ is the likelihood.

Unfortunately, direct inference of the posterior is intractable due to the large parameter space and nonlinear nature of DL architectures. A popular approximation technique, known as VI, formulates the problem of posterior inference as an optimization problem \citep{graves2011practical}. The VI approach considers a simple family of distributions over the network parameters and attempts to find a distribution, called the \emph{variational distribution} $q_{\vc{\theta}}(\mathrm{\Omega})$, within this family that is ``close'' to the \emph{true} unknown posterior. The notion of distributional closeness is captured by the Kullback-Leibler (KL) divergence, and the optimization is performed with respect to the variational distribution parameters $\vc{\theta}$:
\begin{equation}\label{eq:kl}
\mathbf{KL} \left(q_{\vc{\theta}}(\mathrm{\Omega})\big|\big|p(\Omega\big|\mathcal{D})\right)=
\int q_{\vc{\theta}}(\mathrm{\Omega}) \log\frac{  q_{\vc{\theta}}(\mathrm{\Omega})}{p(\mathrm{\Omega}) p( \cal{D} |\mathrm{\Omega})} d \mathrm{\Omega}.
\end{equation}
By rearranging terms in (\ref{eq:kl}), the well-known ELBO objective function is obtained \citep{graves2011practical}:
\begin{equation}\label{eq:elbo}
\mathbf{\mathcal{L}}(\vc{\theta}) = -~\mathbb{E}_{q_{\vc{\theta}}(\Omega)}\left[\log(p(\mathcal{D}|\Omega)\right] + \mathbf{KL} \left(q_{\vc{\theta}}(\mathrm{\Omega})\big|\big|p(\Omega)\right).
\end{equation}

Most Bayesian DL frameworks that use the VI approach sample one set of parameters $\vc{\theta}$ and perform a deterministic forward pass and backpropagation.  The second moment or the variance of the predictive distribution is obtained post-training using MC samples at inference time \citep{blundell2015weight}. This practice is based on the assumption that the single set of sampled parameters $\vc{\theta}$ represents the variational distribution $q_{\vc{\theta}}(\mathrm{\Omega})$ with sufficient accuracy, which has no theoretical grounds \citep{dera2019extended}.
\subsection{Encoder Operations} \label{encoder}
We define a multivariate Gaussian distribution as a prior distribution for all convolution kernels. We assume that kernels are independent within each layer as well as across layers in both the encoder and decoder paths. The independence assumption results in a single additional parameter (variance) for each kernel, limiting the increase in the number of parameters due to the Bayesian formulation. Moreover, independent kernels help extract uncorrelated features and better explore the input space \citep{dera2019extended}.

\noindent
\textbf{Convolution Between Input and Network Parameters:} 
The convolution operation in the first layer is performed between the input data (initially assumed deterministic for simplicity) and the network parameters (random variables). 
We assume that network parameters $\bm{\mathcal{W}}_{e}^{(k_1)}$ follow a Gaussian distribution, i.e., $\text{vec}(\bm{\mathcal{W}}_{e}^{(k_1)})  \sim\mathcal{N}\left(\mathbf{m}_{e}^{(k_1)}, \mathbf{\Sigma}_{e}^{(k_1)}  \right) $.
We write the convolution as a matrix-vector multiplication, where $\mathbf{X}$ denotes the matrix having rows equal to the vectorized sub-tensors of the input $\bm{\mathcal{X}}$.
Then, the convolution operation is expressed as $\bm{\mathbf{z}}_{e}^{(k_1)} = \mathbf{X} \times \text{vec}(\bm{\mathcal{W}}_{e}^{(k_1)}$), for $k_1 = 1, \cdots, K_1$. Thus, the output of the first convolutional layer follows a Gaussian distribution where the mean and covariance are given by: 
\begin{equation}
   \bm{\mathbf{z}}_{e}^{(k_1)} \sim \mathcal{N}\left(\mathbf{X} \mathbf{m}_{e}^{(k_1)},\text{ } \mathbf{X} \mathbf{\Sigma}_{e}^{(k_1)} \mathbf{X}^T \right).  
\end{equation}

\noindent
\textbf{Convolution Between Two Random Variables:}
We consider a generic case of convolution between two random variables. Let $\bm{\mathcal{B}}$ be the incoming input to any convolution layer, except the first layer, i.e., $c \ne 1$. The convolution operation is expressed as a matrix-vector multiplication; however, in this case both the input and the kernels are random tensors.
We form $\mathbf{B}$ by vectorizing the sub-tensors of the incoming input $\bm{\mathcal{B}}$, i.e., $\mathbf{B} = [\mathbf{b}_{1}^T, \mathbf{b}_{2}^T, \cdots, \mathbf{b}_{J}^T]^T$, where $\mathbf{b}_{j}^T$ represents $j^{\text{th}}$ row of $\mathbf{B}$. Let $\bm{\mu}_{\mathbf{b}_{j}}$ and $\mathbf{\Sigma}_{\mathbf{b}_{j}}$ represent the mean and covariance of $\mathbf{b}_{j}$. Then, the output of the convolution is formulated as $\bm{\mathbf{z}}_{e}^{(k_c)} = \mathbf{B}$ $\times$ vec$(\bm{\mathcal{W}}_{e}^{(k_c)})$ with vec$(\bm{\mathcal{W}}_{e}^{(k_c)}) \sim \mathcal{N}\left(\mathbf{m}_{e}^{(k_c)}, \mathbf{\Sigma}_{e}^{(k_c)}  \right)$ for $k_c = 1, \cdots, K_c$ . 
Given that the input $\bm{\mathcal{B}}$ (feature map) is independent from the subsequent layer kernels, we compute elements of the mean of $\bm{\mathbf{z}}_{e}^{(k_c)}$ as the product of the two mean vectors, $\bm{\mu}_{\mathbf{b}_{j}}$ and $\mathbf{m}_{e}^{(k_c)}$, i.e.,
\begin{equation} \label{eq:mu}
[\bm{\mu}_{\bm{\mathbf{z}}_{e}^{(k_c)}}]_j =  \bm{\mu}_{\mathbf{b}_{j}}^T \mathbf{m}_{e}^{(k_c)},~~~j = 1, \cdots, J. 
\end{equation}
\noindent The elements of the covariance matrix $\mathbf{\Sigma}_{\bm{\mathbf{z}}_{e}^{(k_c)}}$ are derived as:
\begin{align}
    & \text{Non-diagonal elements} ~~ (i \neq j): \quad \bm{\mu}_{\mathbf{b}_{i}}^T  \mathbf{\Sigma}_{e}^{(k_c)} \bm{\mu}_{\mathbf{b}_{j}},  \label{eq:random1} \\
    & \text{Diagonal elements} ~~ (i = j): \quad \tr\big( \mathbf{\Sigma}_{\mathbf{b}_{i}}\mathbf{\Sigma}_{e}^{(k_c)}  \big) +\bm{\mu}_{\mathbf{b}_{i}}^T  \mathbf{\Sigma}_{e}^{(k_c)} \bm{\mu}_{\mathbf{b}_{j}}+ \mathbf{m}_{e}^{(k_c)T} \mathbf{\Sigma}_{\mathbf{b}_{j}} \mathbf{m}_{e}^{(k_c)}. \label{eq:random2}
\end{align}


\noindent
\textbf{Nonlinear Activation Function:}
Convolutional layers are commonly followed by an element-wise nonlinear activation function, e.g., Rectified Linear Unit (ReLU). Let $\psi$ denote the activation function and $\mathbf{g}_{e}^{(k_c)}$ denote the output of the activation function, i.e., $\mathbf{g}_{e}^{(k_c)} = \psi[\bm{\mathbf{z}}^{(k_c)}_{e}]$ for $k_c = 1, \cdots, K_c$ . We use the first-order Taylor series approximation to derive the mean and covariance of the random variable $\mathbf{g}_{e}^{(k_c)}$, i.e.,
\begin{align}
    \bm{\mu}_{\mathbf{g}_{e}^{(k_c)}} \approx \psi\big(\bm{\mu}_{\bm{\mathbf{z}}_{e}^{(k_c)}}\big), \quad \bm{\Sigma}_{\mathbf{g}_{e}^{(k_c)}} \approx \mathbf{\Sigma}_{\bm{\mathbf{z}}_{e}^{(k_c)}}\odot \left[ \nabla \psi \big(\bm{\mu}_{\bm{\mathbf{z}}_{e}^{(k_c)}}\big)  \nabla \psi \big(\bm{\mu}_{\bm{\mathbf{z}}_{e}^{(k_c)}}\big) ^T\right],
\end{align}
where $\nabla$ is the gradient with respect to $\bm{\mathbf{z}}^{(k_c)}_{e}$. 

\noindent
\textbf{Max-Pooling Operation:} 
The max-pooling operation is often used to downsample the incoming feature map. 
We propagate the mean through the max-pooling layer using the classical operation of selecting the largest value from a patch in the feature map. The pooling for the covariance is achieved by only retaining the rows and columns (of the incoming covariance matrix) corresponding to the retained elements (pooled elements) of the mean vector. We write the mean and covariance as follows:
\begin{align}
\bm{\mu}_{\mathbf{p}_{e}^{(k_c)}} = \text{pool}(\bm{\mu}_{\mathbf{g}_{e}^{(k_c)}}), \quad \mathbf{\Sigma}_{\mathbf{p}_{e}^{(k_c)}} = \text{co-pool}(\mathbf{\Sigma}_{\mathbf{g}_{e}^{(k_c)}}). 
\end{align}
An encoder may consist of multiple layers of convolution operations, nonlinear activation functions, and max-pooling to get a low-dimensional representation of the input.

\subsection{Decoder Operations} \label{decoder}
The operations in the decoder path start with the low-dimensional representation produced by the encoder. The decoder may also include multiple convolutional layers, which are performed following the mathematical relationships provided in Eqs. (\ref{eq:mu})-(\ref{eq:random2}).

\noindent
\textbf{Up-sampling:}
The up-sampling is an essential part of the decoder path that increases the resolution of the input. Using $\mathbf{g}_{d}^{(k_c)}$ to represent the input to the up-sampling operation and ${\mathbf{u}_{d}^{(k_c)}} $ as the output, we have:
\begin{align}
    \mathbf{u}_{d}^{(k_c)} = \text{up-sample}\left({\mathbf{g}_{d}^{(k_c)}}\right).    
\end{align}
The mean of $\mathbf{u}_{d}^{(k_c)}$ is computed by inserting zeros between two consecutive elements of the input and padding with zeros. The covariance matrix is obtained by adding rows and columns of zeros at locations corresponding to the newly added zeros in the mean. 

\noindent
\textbf{Up-convolution:}
The up-sampling operation may produce sparse feature maps with many zeros. Generally, a $2 \times 2$ convolution operation is performed to get a dense high-resolution output. The mean and covariance are computed using results presented in Eqs. (\ref{eq:mu})-(\ref{eq:random2}).

\noindent
\textbf{Padding:} 
The padding operation applied to the mean is the same as the classical zero-padding operation. For the covariance matrix, we add a new row and a new column for each element padded to the mean. The new elements added in the covariance matrix are all set to zero, and the variance (diagonal) elements are set to a user-defined small value with $\sigma_{pa} > 0$.

\noindent
\textbf{Concatenation:} 
The features from the encoder side are generally concatenated with the corresponding features from the decoder to improve the localization of various objects in the input. The feature maps from the encoder path may need to be resized or cropped before they can be concatenated with the decoder features due to the differences in size. 

Let $\bm{\mathcal{G}}_{e}^{c}$ be the $c^\text{th}$ encoder feature map, and $\mathbf{g}_{e}^{(k_c)}$ the $k_c^\text{th}$ slice from such map with mean and covariance $\bm{\mu}_{\mathbf{g}_{e}^{(k_c)}}$ and $ \mathbf{\Sigma}_{\mathbf{g}_{e}^{(k_c)}}$, respectively.  The cropped feature map is denoted with $\bm{\mathcal{G}}_{e}^{*c} $ where $k_c^\text{th}$ slice is $\mathbf{g}_{e}^{*(k_c)}$. For $k_c = 1, \ldots, K_c$, $\bm{\mu}_{\mathbf{g}_{e}^{*(k_c)}}=\text{crop}(\bm{\mu}_{\mathbf{g}_{e}^{(k_c)}})$ while $\mathbf{\Sigma}_{\mathbf{g}_{e}^{*(k_c)}}$ is obtained by removing the rows and columns from $\mathbf{\Sigma}_{\mathbf{g}_{e}^{(k_c)}}$ corresponding to the cropped elements of $\bm{\mu}_{\mathbf{g}_{e}^{(k_c)}}$.  

The output of the concatenation operation is a feature map $\bm{\mathcal{G}}_{d}^{*c} = \{ \bm{\mathcal{G}}_{d}^{c},\bm{\mathcal{G}}_{e}^{*c} \}$, where $\bm{\mathcal{G}}_{d}^{c}$ is the $c^\text{th}$ decoder feature map.
The concatenation operation is done along the dimension that represents channels in the feature maps (generally the third dimension).

\noindent
\textbf{Softmax Function:}
Pixel-level segmentation can be considered as a dense classification problem where we assign a label to each pixel. Hence, for a multi-class problem, a softmax function $\phi$ is applied to the output of the last layer.
Let $\mathbf{F}$ represent the output of the last layer with mean $\bm{\mu}_{\mathbf{F}}$ and covariance $\bm{\Sigma}_{\mathbf{F}}$, and $\mathbf{Y}$ denote the output of the network after the softmax operation. We can approximate the mean $\bm{\mu}_{\mathbf{Y}}$ and covariance $\bm{\Sigma}_{\mathbf{Y}}$ using first-order Taylor series, that is:
\begin{equation}
    \begin{aligned}
        \bm{\mu}_{\mathbf{Y}} & \approx \phi(\bm{\mu}_{\mathbf{F}}), \quad
        \bm{\Sigma}_{\mathbf{Y}}  \approx \bm{J}_{\phi}\bm{\Sigma}_{\mathbf{F}}\bm{J}_{\phi}^T,
    \end{aligned}
\end{equation}
where $\bm{J}_{\phi}$ is the Jacobian matrix of $\phi$ computed with respect to $\mathbf{F}$ evaluated at $\bm{\mu}_{\mathbf{F}}$.

The mathematical results presented above for various operations can be used to build any type of deep NN in addition to the proposed encoder-decoder-based networks. 

\section{Experimental Methods}\label{set up}

\subsection{Datasets}
We use three different medical benchmark segmentation datasets, including lung CT \citep{ma_jun_2020_3757476}, hippocampus MRIs \citep{antonelli2022medical} and brain tumor MRIs \citep{ menze2014multimodal}, and one clinical dataset \citep{pmid31136584}. Our experiments use only the publicly available annotated data from the respective datasets, i.e., unlabeled data is not used for training or validation. The datasets are divided into training, validation and testing bins with approximately $80\%$ selected for training, $10\%$ for validation and $10\%$ for testing. 

\subsubsection{Lungs Dataset} 
The dataset includes $20$ CT scans from the chest region, available at \href{https://zenodo.org/records/3757476}{zenodo.org}  \citep{ma_jun_2020_3757476}. This heterogeneous dataset consists of both COVID-19 and non-COVID-19 patients. The data annotations include left lung, right lung and infections (if found). We consider a binary segmentation task for this dataset, i.e., delineating the boundaries of the lungs in the given CT images. We assign a label of $0$ to the background and $1$ to lung tissue. The pre-processing steps include: 1) windowing the Hounsfield units range between $-1250$ and $250 $; 2) normalizing all pixel values between $0$ and $1$; 3) deleting empty slices, i.e., slices that include only the label $0$ corresponding to the background to minimize class imbalance; and 4) cropping all images to a single size, i.e.,  $ 512 \times 512$ pixels.
 
\subsubsection{Hippocampus Dataset}
The Hippocampus data is available as part of the \href{http://medicaldecathlon.com/}{Medical Segmentation Decathlon} \citep{antonelli2022medical}. The dataset consists of 394 single-modality MRI scans. The segmentation task requires the precise delineation of two adjacent structures, i.e., anterior (label $1$) and posterior (label $2$). The pre-processing steps include: 1) normalizing the data to reduce the image bias (which is a characteristic of MRI data); 2) deleting empty slices, i.e., those that include only the background; and 3) padding images to have the same input size of $64 \times 64$ pixels.  

\subsubsection{Brain Tumor Segmentation (BraTS) Dataset}
The Brain Tumor Segmentation (BraTS) dataset is available as part of the \href{http://braintumorsegmentation.org/}{MICCAI BraTS Challenge}. The dataset includes about 300 multi-modal (T1, T1c, T2, and FLAIR) MRI scans from $274$ brain tumor patients (some patients have multiple MRI scans) \citep{menze2014multimodal}. The dataset is divided into two main types of tumors: low-grade gliomas (LGG) and high-grade gliomas (HGG). We focus on the more challenging HGG dataset in our experiments. The pre-processing steps include: 1) normalizing data to reduce the image bias; 2) deleting images that do not include any tumor structure; and 3) cropping each image to the size of $240 \times 240$ pixels. The input data size for each sample in the dataset is $240 \times 240 \times 4$ pixels, where the last number represents the four modalities, i.e., T1, T1c, T2, and FLAIR.  
All four networks (U-Net, Bayes U-Net, and SUPER U-Net) are trained to segment $5$ different labels in the HGG MRIs, i.e., normal tissue (label $0$), necrosis (label $1$), edema (label $2$),  non-enhancing tumor (label  $3$), and enhancing tumor (label $4$). In most clinical applications, generally, three tumor regions are considered for evaluating the results of segmentation: whole tumor (labels 1, 2, 3 and 4), tumor core (labels 1, 3 and 4), and enhancing tumor region (label 4) \citep{menze2014multimodal}. 

\subsubsection{Clinical Dataset} We acquired a real-world, anonymized, IRB-approved brain tumor dataset from the O’Neal Comprehensive Cancer Center at the University of Alabama at Birmingham (UAB) School of Medicine. This dataset will be made available upon request. The imaging dataset includes $627$ fluid-attenuated inversion recovery (FLAIR) sequences, including 24 images each on average, from patients diagnosed with World Health Organization grade 2 gliomas, seen at the neuro-oncology clinics at the University of Alabama at Birmingham \citep{pmid31136584}. The tumor masks were manually annotated by an expert physician. The pre-processing steps include: 1) normalizing data to reduce the image bias; 2) deleting images that do not include any tumor structure; and 3) cropping each image to the size of $240 \times 240$ pixels. 

\subsection{Segmentation Network Architectures}
\begin{table}[t]
\footnotesize
\centering
\caption{Architecture details and training hyperparameters for different datasets.}
\begin{tabular}{c|c|c|c|c|c|c|c}
    \hline
    \textbf{Dataset} & \textbf{\begin{tabular}[c]{@{}c@{}}Encoder\\ Blocks\end{tabular}} & \textbf{\begin{tabular}[c]{@{}c@{}}Decoder\\ Blocks\end{tabular}} & \textbf{\begin{tabular}[c]{@{}c@{}}Encoder\\ Filters\end{tabular}} & \textbf{\begin{tabular}[c]{@{}c@{}}Decoder\\ Filters\end{tabular}} & \textbf{\begin{tabular}[c]{@{}c@{}}Batch\\ size\end{tabular}} & \textbf{Epochs} & \begin{tabular}[c]{@{}c@{}}\boldmath{$\sigma_{pa}$}\\ \begin{tabular}[c]{@{}c@{}}SUPER\\ U-Net\end{tabular}\end{tabular}  \\
    \hline
    Lungs        & 3 & 2 & 16, 32, 64 & 32, 16 & 10 & 50  & 0.1  \\ \hline
    Hippocampus  & 3 & 2 & 32, 64, 128 & 64, 32 & 20 & 100 & 0.02 \\ \hline
    BraTS        & 5 & 4 & \begin{tabular}[c]{@{}c@{}}64, 128, 256,\\ 512, 1024\end{tabular}  &  \begin{tabular}[c]{@{}c@{}}512, 256,\\ 128, 64\end{tabular} & 20 & 100 & 0.1  \\ \hline
    Clinical     & 5 & 4 & \begin{tabular}[c]{@{}c@{}}16, 32, 64,\\ 128, 256\end{tabular}  &  \begin{tabular}[c]{@{}c@{}}128, 64,\\ 32, 16\end{tabular} & 10 & 100 & 0.01 \\ \hline
    \multicolumn{8}{l}{All experiments used the Adam optimizer with a learning rate of 0.001.
} \\

\end{tabular}
\label{tab:architecture_hyperparams}

\end{table}

We apply the proposed SUPER-Net framework to the U-Net architecture; for simplicity, we refer to it as SUPER U-Net. We compare SUPER U-Net with three state-of-the-art segmentation networks, a deterministic U-Net \citep{ronneberger2015u}, a Bayes U-Net obtained using MC dropout \citep{kendall2017bayesian}, and an Ensemble U-Net \citep{lakshminarayanan2017simple} 
\subsubsection{U-Net - The Baseline Segmentation Architecture}
Among all architectures proposed for medical image segmentation, U-Net is the most widely used \citep{ronneberger2015u}. 
U-Net is built using the encoder-decoder structure with a contracting path that is almost identical to the expanding path. The contracting path may consist of multiple encoder blocks, which, in turn, may include various convolution layers, max-pooling, and nonlinear activations. The expanding path consists of multiple decoder blocks, which are made of multiple layers of convolution, activation functions, up-convolution, up-sampling and padding. Additionally, there are connections between the encoder and decoder blocks that concatenate feature maps from the encoder with the corresponding feature maps of the decoder. Finally, a $1 \times 1$ convolution and SoftMax are applied to the decoded feature maps before calculating the cross-entropy loss function.

In the original U-Net architecture \citep{ronneberger2015u}, the border pixels are lost due to un-padded convolution operations and the missing regions are extrapolated by mirroring. Such processing may yield erroneous results for some medical image segmentation datasets. Hence, in our setting, we apply the padding operation to increase the size of the feature maps and reconstruct the full image at the output of the network. We include the padding operation twice in each decoder block on the expanding path. The first padding operation is performed before the concatenation, and the second is performed before the second convolution in each decoder block. In our experiments, we refer to this U-Net architecture as the deterministic segmentation network.

In Table \ref{tab:architecture_hyperparams}, we report the specifics of the architectures for all datasets. The kernel size is set to $3$ for all datasets. For clinical data, we use convolutions with padding set to \emph{same}, and apply batch normalization on both the encoder and decoder.

\subsubsection{Bayes U-Net}
Bayes U-Net is built using the MC dropout technique following the implementation of \citep{kendall2017bayesian}. The dropout is used only in the central blocks with the probability of dropping a neuron set to $p=0.5$. Bayes U-Net uses cross-entropy loss function. At the inference time, we use $N = 20$ MC samples and the uncertainty is measured in terms of predictive entropy (PE) \citep{gal2016dropout}.

\subsubsection{Ensemble U-Net}
Ensemble U-Net is built using an ensemble of U-Net models. We trained $5$ networks with different initializations and used the entire training set for each model. The number of networks is chosen following the results in \citep{ larrazabal2021orthogonal}. Ensemble U-Net uses the cross-entropy loss function. At inference time, each input is fed to the five models and the outputs are used to estimate uncertainty using PE.

\subsubsection{SUPER U-Net}
SUPER U-Net uses the mathematical operations presented in Sections \ref{encoder} and \ref{decoder} to propagate the first two moments of the variational distribution through the U-Net architecture. The output of SUPER U-Net consists of a segmentation map and an uncertainty map. The former is given by the mean of the predictive distribution, while the latter is generated by the covariance of the predictive distribution. We use a Gaussian variational distribution and employ the ELBO loss function defined in Eq. (\ref{eq:elbo}). We optimize the ELBO loss function with respect to the variational parameters, i.e., the mean and covariance of the variational distribution. 
To reduce the computational complexity, we propagate diagonal covariance matrices.

\subsection{Other Experimental Settings}
We report the specific hyperparameters used for each dataset in Table \ref{tab:architecture_hyperparams}, including the optimizer, learning rate, batch size, number of training epochs, and $\sigma_{pa}$ for the padding in the SUPER U-Net model. The selection was determined through empirical evaluation. We explored different values for each hyperparameter and selected those that provided stable training, faster convergence, and improved segmentation performance across all datasets. The values of $\sigma_{pa}$ in SUPER U-Net were tuned to balance the trade-off between predictive uncertainty and segmentation accuracy. The batch size was determined based on the size of the data and the available hardware constraints. All simulations were performed using Python with the TensorFlow library on an NVIDIA RTX A6000 GPU.

We report the Dice Similarity Coefficient (DSC) as the metric to compare the performance of all four networks. 
We conduct a detailed robustness analysis of the performance of all four networks using two types of noise, i.e., Gaussian and adversarial. We compare the performance of all four networks under various levels of Gaussian noise added to the test data of all three datasets. We measure the noise level using the signal-to-noise ratio (SNR) in the units of decibels (dB). For the adversarial noise, we use the Fast Gradient Sign Method (FSGM) to generate untargeted attacks \citep{liu2017delving}, and we use the Projected Gradient Descent (PGD) method to generate targeted adversarial attacks \citep{madry2018towards}. The attacks are generated with a maximum number of iterations set to $20$ and a step size of $1$. We select a \emph{source} class and a \emph{target} class to generate targeted attacks. The adversarial attack algorithm will try to fool the trained network into predicting pixels belonging to the \emph{source} class as the pixels of the \emph{target} class.

\section{Results}\label{results}
We report our results in four parts. First, we present the performance analysis (measured using DSC) of the four networks (U-Net, Bayes U-Net, Ensemble U-Net, and SUPER U-Net) under various levels of Gaussian noise added to the benchmark test datasets. Next, we analyze the same four networks under various levels of targeted and untargeted adversarial attacks. We report the results of the clinical data. Finally, we present an analysis of the uncertainty maps and the predictive variance generated by the proposed SUPER U-Net at inference time. For reference, we report DSC values for U-Net, Bayes U-Net, Ensemble U-Net and SUPER U-Net for noise-free test data in tables \ref{tab:gaussian_lung},  \ref{tab:gaussian_hip}, and \ref{tab:gaussian}.
\begin{table}
\footnotesize
\centering
\caption{DSC for Lungs Dataset - performance comparison under additive Gaussian noise.}
\label{tab:gaussian_lung}
\begin{tabular}{lcccc}
 \hline \hline
\multicolumn{1}{l|}{}                     & \multicolumn{1}{c|}{\textbf{U-Net}} &\multicolumn{1}{c|}{\textbf{Bayes U-Net}} & \multicolumn{1}{c|}{\textbf{Ensemble U-Net}} & \multicolumn{1}{c}{\textbf{SUPER U-Net}} \\ \hline \hline
\multicolumn{1}{c|}{Noise Free} & \multicolumn{1}{c|}{\textbf{.83}} &\multicolumn{1}{c|}{\textbf{.83}}          & \multicolumn{1}{c|}{\textbf{.83}}        & \multicolumn{1}{c}{\textbf{.83}}       \\ \hline
\multicolumn{5}{c}{Gaussian noise added  to the entire image}                                                                                           \\ \hline
\multicolumn{1}{l|}{SNR $\approx 35$ dB}     & \multicolumn{1}{c|}{.82}       &\multicolumn{1}{c|}{.82}      & \multicolumn{1}{c|}{.82}        & \multicolumn{1}{c}{\textbf{.83}}       \\ \hline
\multicolumn{1}{l|}{SNR $\approx 3$ dB}        & \multicolumn{1}{c|}{.16}               & \multicolumn{1}{c|}{.19}   & \multicolumn{1}{c|}{.11}          & \multicolumn{1}{c}{\textbf{.21}}           \\ \hline
\multicolumn{5}{c}{Gaussian noise added to lung pixels only}                                                                                             \\ \hline
\multicolumn{1}{l|}{SNR $\approx 31$ dB}     & \multicolumn{1}{c|}{.82}           & \multicolumn{1}{c|}{\textbf{.83}}        & \multicolumn{1}{c|}{\textbf{.83}} &  \multicolumn{1}{c}{\textbf{.83}}      \\ \hline
\multicolumn{1}{l|}{SNR $\approx 14$ dB}        & \multicolumn{1}{c|}{.63}   & \multicolumn{1}{c|}{.63}      & \multicolumn{1}{c|}{.65}            & \multicolumn{1}{c}{\textbf{.79}}            \\ \hline

\end{tabular}
\end{table}

\begin{table*}[t]
\centering
\caption{DSC for Hippocampus Dataset - Noise Free.}
\label{tab:gaussian_hip}
\footnotesize
\begin{tabular}{ccccccccc}
 \hline \hline
 \multicolumn{1}{l}{}                     & \multicolumn{4}{|c|}{{\textbf{Anterior}}}                                                                                                                                                                 & \multicolumn{4}{c}{\textbf{Posterior}}                                                                                                                                                               \\ 
  \hline \hline 
\multicolumn{1}{l|}{}                     & \multicolumn{1}{c|}{\textbf{U-Net}} & \multicolumn{1}{c|}{
\textbf{\begin{tabular}[c]{@{}c@{}}Bayes\\ U-Net\end{tabular}}} & \multicolumn{1}{c|}{\textbf{\begin{tabular}[c]{@{}c@{}}Ensemble\\ U-Net\end{tabular}}} & \multicolumn{1}{c|}{\textbf{\begin{tabular}[c]{@{}c@{}}SUPER\\ U-Net\end{tabular}}} & \multicolumn{1}{c|}{\textbf{U-Net}} & \multicolumn{1}{c|}{\textbf{\begin{tabular}[c]{@{}c@{}}Bayes\\ U-Net\end{tabular}}} & \multicolumn{1}{c|}{\textbf{\begin{tabular}[c]{@{}c@{}}Ensemble\\ U-Net\end{tabular}}} &\multicolumn{1}{c}{\textbf{\begin{tabular}[c]{@{}c@{}}SUPER\\ U-Net\end{tabular}}} \\ 

\hline \hline
\multicolumn{1}{l|}{Noise Free} & \multicolumn{1}{c|}{\textbf{.79}}  & \multicolumn{1}{c|}{\textbf{.79}}                                                   & \multicolumn{1}{c|}{\textbf{.79}} & \multicolumn{1}{c|}{\textbf{.79}}                                                 & \multicolumn{1}{c|}{.76}  & \multicolumn{1}{c|}{.76}        & \multicolumn{1}{c|}{\textbf{.77}}                                             & \multicolumn{1}{c}{.74}                                                          \\ \hline

\end{tabular}
\end{table*}


\begin{table*}
\centering
\caption{DSC for BraTS Dataset - Noise Free}
\label{tab:gaussian}
\footnotesize
\resizebox{\textwidth}{!}{
\begin{tabular}{l|c|c|c|c|c|c|c|c|c|c|c|c|}
\hline \hline
   & \multicolumn{4}{c|}{\textit{\textbf{Whole}}}    & \multicolumn{4}{c|}{\textit{\textbf{Core}}}           & \multicolumn{4}{c}{\textit{\textbf{Enhancing}}}    \\ 
                                          \hline \hline
& \multicolumn{1}{c|}{\textbf{U-Net}} & \multicolumn{1}{c|}{\textbf{\begin{tabular}[c]{@{}c@{}}Bayes\\ U-Net\end{tabular}}} & \multicolumn{1}{c|}{\textbf{\begin{tabular}[c]{@{}c@{}}Ensemble\\ U-Net\end{tabular}}} & \multicolumn{1}{c|}{\textbf{\begin{tabular}[c]{@{}c@{}}SUPER\\ U-net\end{tabular}}} & \multicolumn{1}{c|}{\textbf{U-Net}} & \multicolumn{1}{c|}{\textbf{\begin{tabular}[c]{@{}c@{}}Bayes\\ U-Net\end{tabular}}} & \multicolumn{1}{c|}{\textbf{\begin{tabular}[c]{@{}c@{}}Ensemble\\ U-Net\end{tabular}}} & \multicolumn{1}{c|}{\textbf{\begin{tabular}[c]{@{}c@{}}SUPER\\ U-net\end{tabular}}} & \multicolumn{1}{c|}{\textbf{U-Net}} & \multicolumn{1}{c|}{\textbf{\begin{tabular}[c]{@{}c@{}}Bayes\\ U-Net\end{tabular}}} & \multicolumn{1}{c|}{\textbf{\begin{tabular}[c]{@{}c@{}}Ensemble\\ U-Net\end{tabular}}} & \multicolumn{1}{c}{\textbf{\begin{tabular}[c]{@{}c@{}}SUPER\\ U-net\end{tabular}}} \\  \hline \hline

\multicolumn{1}{l|}{\begin{tabular}[c]{@{}c@{}}Noise\\ Free\end{tabular}} & .77                                       & .77           & .76                          & \textbf{.83}                          & .58                                       & .58    & .60                                   & \textbf{.64}                          & .57                                       & .57     & .63                                    & \multicolumn{1}{c}{\textbf{.69}}                          \\ \hline

\end{tabular}}
\end{table*}

\begin{figure*}
\begin{subfigure}{0.30\linewidth}
\includegraphics[width=\linewidth]{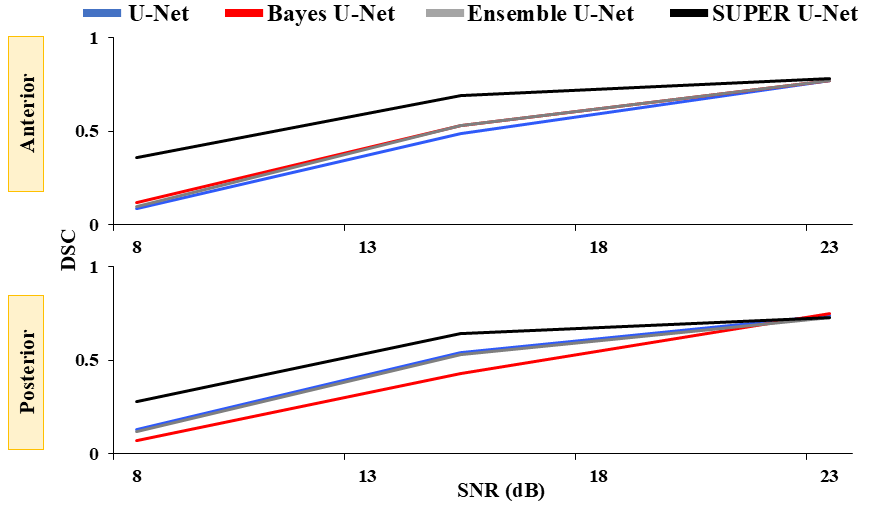}
\caption{Noise added to entire image}
\end{subfigure}
\hfill
\begin{subfigure}{0.30\linewidth}
\includegraphics[width=\linewidth]{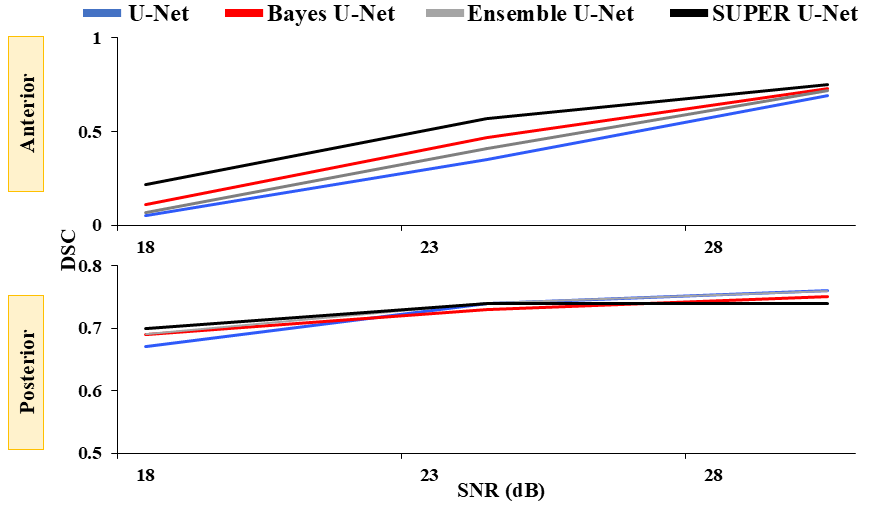}
\caption{Noise added to Anterior pixels}
\end{subfigure}
\hfill
\begin{subfigure}{0.30\linewidth}
\includegraphics[width=\linewidth]{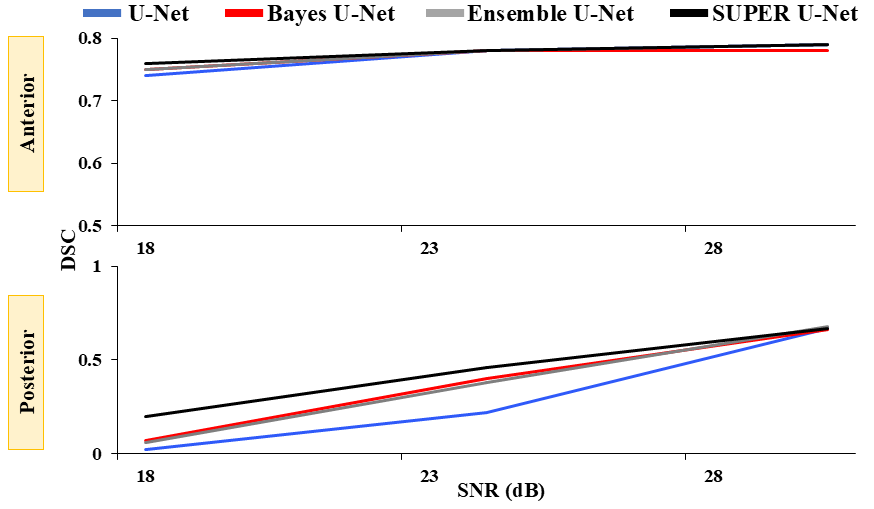}
\caption{Noise added to Posterior pixels}
\end{subfigure}
\caption{Performance of the four networks, i.e., U-Net (blue), Bayes U-Net (red), Ensemble U-Net (gray), and SUPER U-Net (black), under various levels of Gaussian noise added to the (a) entire image, (b) Anterior pixels only, and (c) Posterior pixels only of the Hippocampus test data. We plot Dice Similarity Coefficient (DSC) versus Signal to Noise Ratios (SNRs) for the Anterior and Posterior hippocampus.}
\label{fig:hippo_gaussian_dsc}
\end{figure*}
\begin{figure*}[t]
\centering
    \begin{subfigure}[b]{0.98\textwidth}
        \centering
        \includegraphics[width=\textwidth]{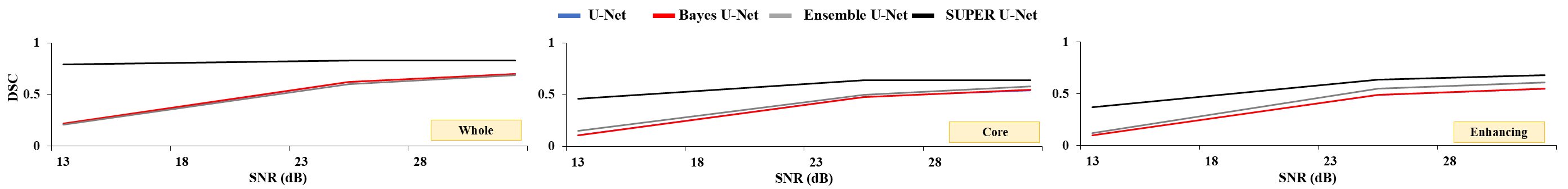}
        \caption{Gaussian noise added to the tumor pixels only}
        \label{fig:brats_gaussian_1_dsc}
    \end{subfigure}
    
    
    \begin{subfigure}[b]{0.98\textwidth}
        \centering
        \includegraphics[width=\textwidth]{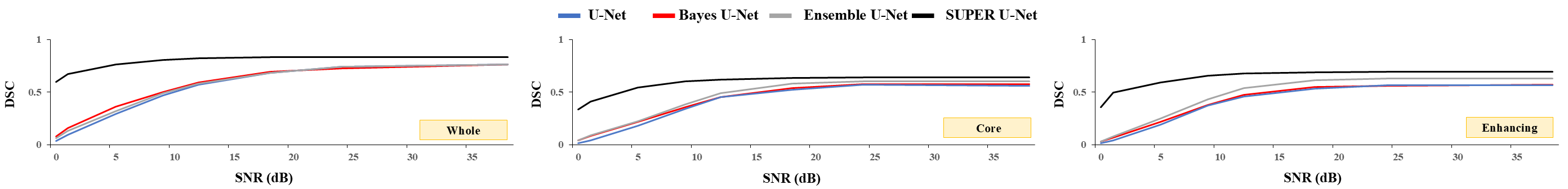}
        \caption{Gaussian noise added to all pixels}
        \label{fig:brats_gaussian_2_dsc}
    \end{subfigure}
    
    \caption{Performance of the four networks, i.e., U-Net (blue),
Bayes U-Net (red), Ensemble U-Net (gray) and SUPER U-Net (black), under various levels of Gaussian noise added to (a) the tumor pixels only and (b) all pixels of the BraTS test data. The three sub-plots show the Dice Similarity Coefficient (DSC) values for a range of Signal to Noise Ratios (SNRs) for three different tumor regions: whole tumor, core, and enhancing.}
    \label{fig:combined_figures}
\end{figure*}


%
\begin{figure}
\centering
\hspace{0cm}\includegraphics[width=0.40\textwidth]{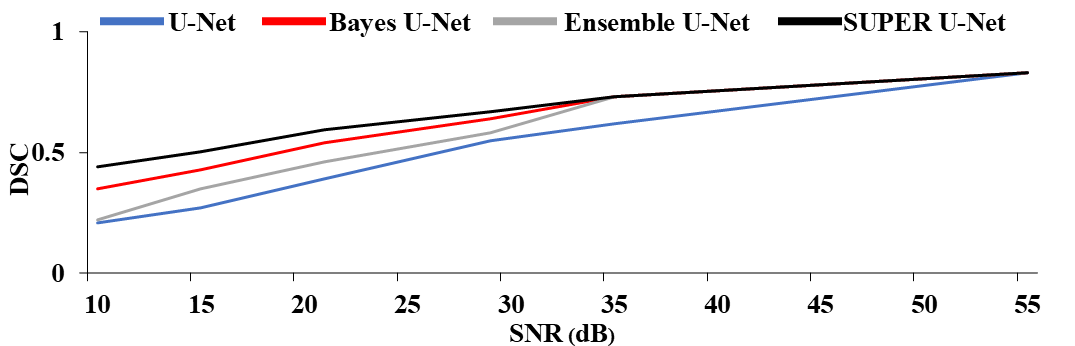}
\caption{Performance of four networks, i.e., U-Net (blue),
Bayes U-Net (red), Ensemble U-Net (gray) and SUPER U-Net (black), under various levels of untargeted attacks to the Lungs test data. We display Dice Similarity Coefficient (DSC) values for a range of Signal to Noise Ratio (SNR).  } \label{fig:lungs_adv_dsc}
\end{figure}

\begin{figure*}
\begin{subfigure}{0.45\linewidth} 
\includegraphics[width=\linewidth]{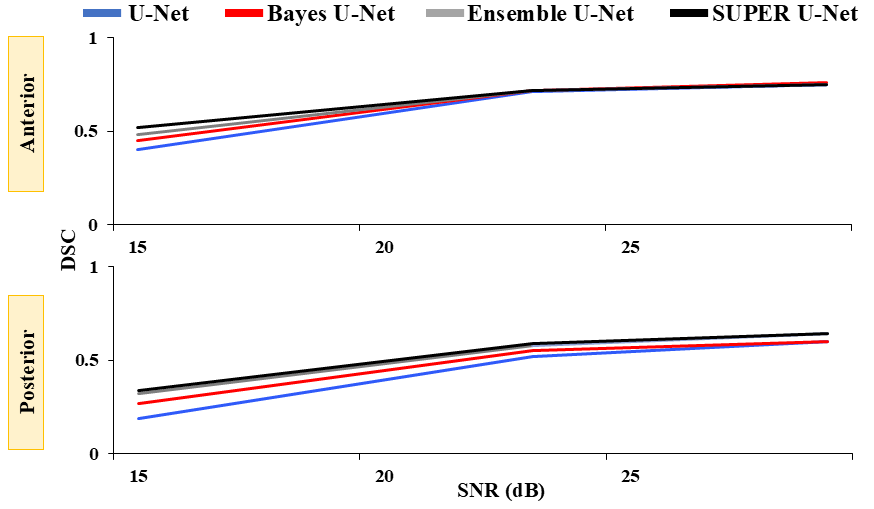}
\caption{Targeted attacks - source: label 1, target: label 2}
\end{subfigure}
\hfill 
\begin{subfigure}{0.45\linewidth} 
\vspace{10pt} 
\includegraphics[width=\linewidth]{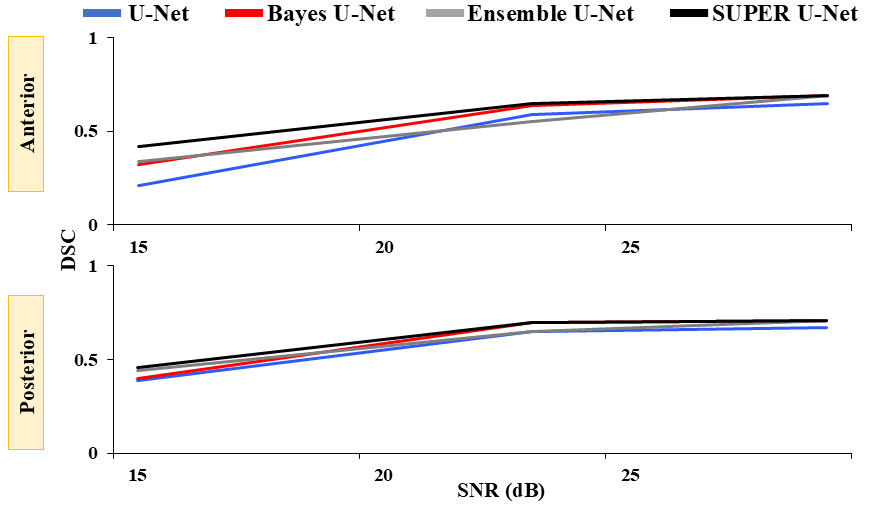}
\caption{Targeted attacks - source: label 2, target: label 1}
\end{subfigure}

\caption{Performance of the four networks, i.e., U-Net (blue), Bayes U-Net (red), Ensemble U-Net (gray), and SUPER U-Net (black), under various levels of adversarial attacks applied to the Hippocampus test data. 
We show targeted adversarial attacks with (a) source:
label 1, target: label 2, (b) viceversa. 
The two subplots show the Dice Similarity Coefficient (DSC) values for the Anterior and Posterior hippocampus measured using Signal to Noise Ratios (SNRs).}
\label{fig:hippo_adv_dsc}
\end{figure*}


\subsection{Evaluation Under Gaussian Noise}
Table \ref{tab:gaussian_lung}, and Figs. \ref{fig:hippo_gaussian_dsc} and \ref{fig:combined_figures} show DSC values for U-Net, Bayes U-Net, Ensemble U-Net and SUPER U-Net under different levels of Gaussian noise. For each dataset, we report results for two cases, i.e., noise added to the entire input image or only to the structures that the networks are trying to segment, e.g., tumors in the BraTS dataset. 

In Table \ref{tab:gaussian_lung}, we compare the performance of the four models for the noise-free test data and for two levels of Gaussian noise added to the entire image and the lung pixels only. Fig. \ref{fig:hippo_gaussian_dsc}, reports the performance of the four models when Gaussian noise is applied to the Hippocampus test data. We consider $3$ scenarios: noise added to the entire image, the Anterior pixels only, and the Posterior pixels only.
We show the results for the BraTS test data in Fig. \ref{fig:combined_figures}. We plot DSCs vs. SNR for the three tumor regions. Each subplot compares the performance of the four networks for multiple levels of Gaussian noise added to the tumor pixels only (Fig. \ref{fig:brats_gaussian_1_dsc}) and the entire image (Fig. \ref{fig:brats_gaussian_2_dsc}). 
The proposed SUPER U-Net generally demonstrates more robust behavior as compared to other models especially at low SNR values, i.e., high levels of noise. 
\subsection{Evaluation Under Adversarial Attacks}
\begin{figure}
\centering
\includegraphics[width=1\textwidth]{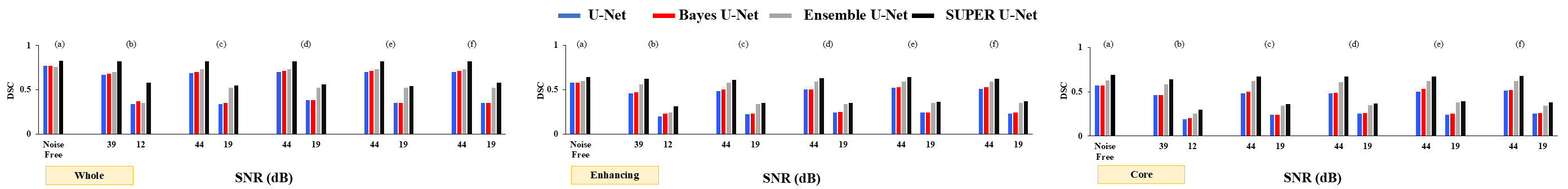}
\caption{Performance of the four networks, i.e., U-Net (blue),
Bayes U-Net (red), Ensemble U-Net (gray) and SUPER U-Net (black) under various levels of adversarial attacks applied to the BraTS test data. The three sub-plots show the Dice Similarity Coefficient (DSC) values for a range of Signal-to-Noise Ratios (SNRs) for three different tumor regions: whole, core, and enhancing tumor. We show (a) noise-free case, (b) untargeted attacks generated using FGSM, (c) Targeted adversarial attacks with source: label 3, target: label 1, (d) Targeted adversarial attacks with source: label 1, target: label 3, (e) Targeted adversarial attacks with source: label 3, target: label 2,  and (f) Targeted adversarial attacks with source: label 2, target: label 3.} \label{fig:brats_adv_dsc}
\end{figure}

We assess the robustness of all four networks against targeted and untargeted adversarial attacks. We show the results in Figures \ref{fig:lungs_adv_dsc}, \ref{fig:hippo_adv_dsc}, and \ref{fig:brats_adv_dsc}. We plot the DSC vs. SNR for the four approaches.

In Fig. \ref{fig:lungs_adv_dsc}, we show DSC values for a range of untargeted adversarial attacks generated using FGSM against the lung test dataset. In Fig. \ref{fig:hippo_adv_dsc}, we consider various levels of targeted attacks applied to (a) the Anterior pixels only and (b) the Posterior pixels only of the Hippocampus test data. For both attack types, we report the performance for the two structures of interest, i.e., anterior and posterior hippocampus. On the other hand, Fig. \ref{fig:brats_adv_dsc} presents both targeted and untargeted adversarial attacks applied to the BraTS test data. The three subplots compare the performance of the four networks on the three structures of interest: whole tumor, core and enhancing tumor. We observe that SUPER U-Net shows better performance (i.e., high DSC values) as compared to the other three networks, especially for stronger attacks (i.e., low values of SNR). 

\begin{figure*}
\begin{center}
\hspace{0cm}\includegraphics[width=1\textwidth]{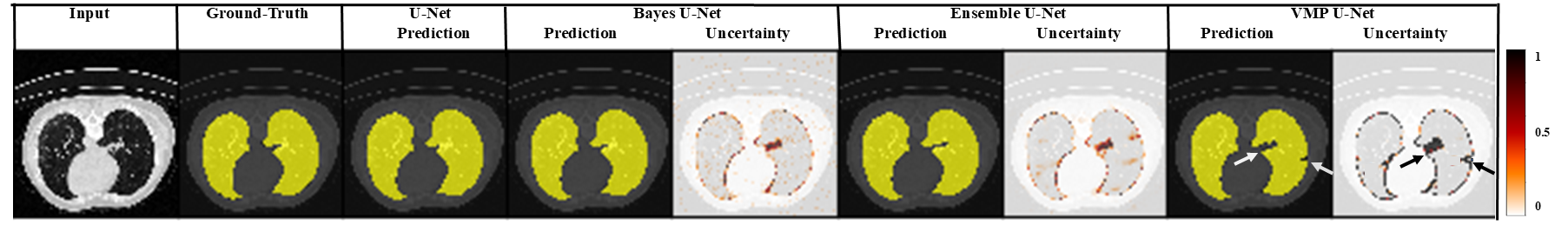}
\caption{Segmentation of the Lungs test data. We show (left to right) the CT input image, the ground-truth segmentation, noise-free models' segmentation predictions (U-Net, Bayes U-Net, Ensemble U-Net, and SUPER U-Net). The uncertainty map of each Bayesian model is shown next to the corresponding prediction. The white arrows point to regions incorrectly classified by the network. We note that the corresponding pixels in the uncertainty maps reflect the low confidence by responding with higher variance values.}\label{fig:lungs_images}
\end{center}
\end{figure*}

\begin{figure*}
\centering
    \begin{subfigure}[b]{0.98\textwidth}
        \centering
        \includegraphics[width=\textwidth]{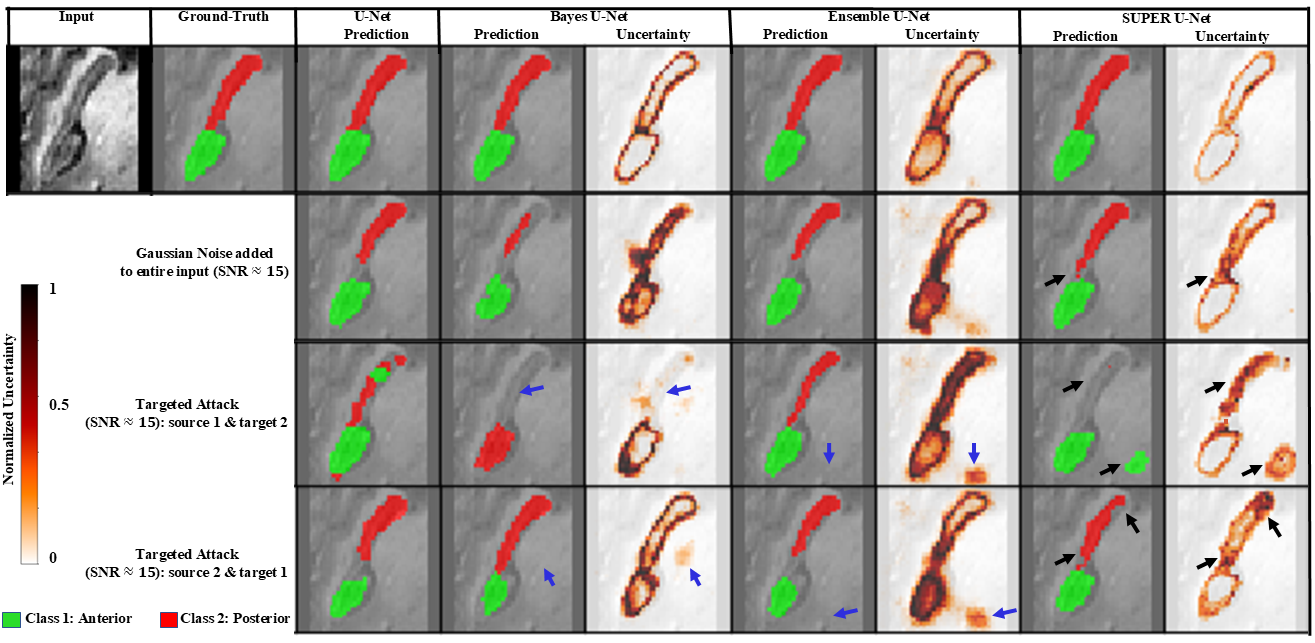}
        \caption{Hippocampus test data}
        \label{fig:hipp}
    \end{subfigure}
    
    \vspace{0.5cm} 
    
    \begin{subfigure}[b]{0.98\textwidth}
        \centering
        \includegraphics[width=\textwidth]{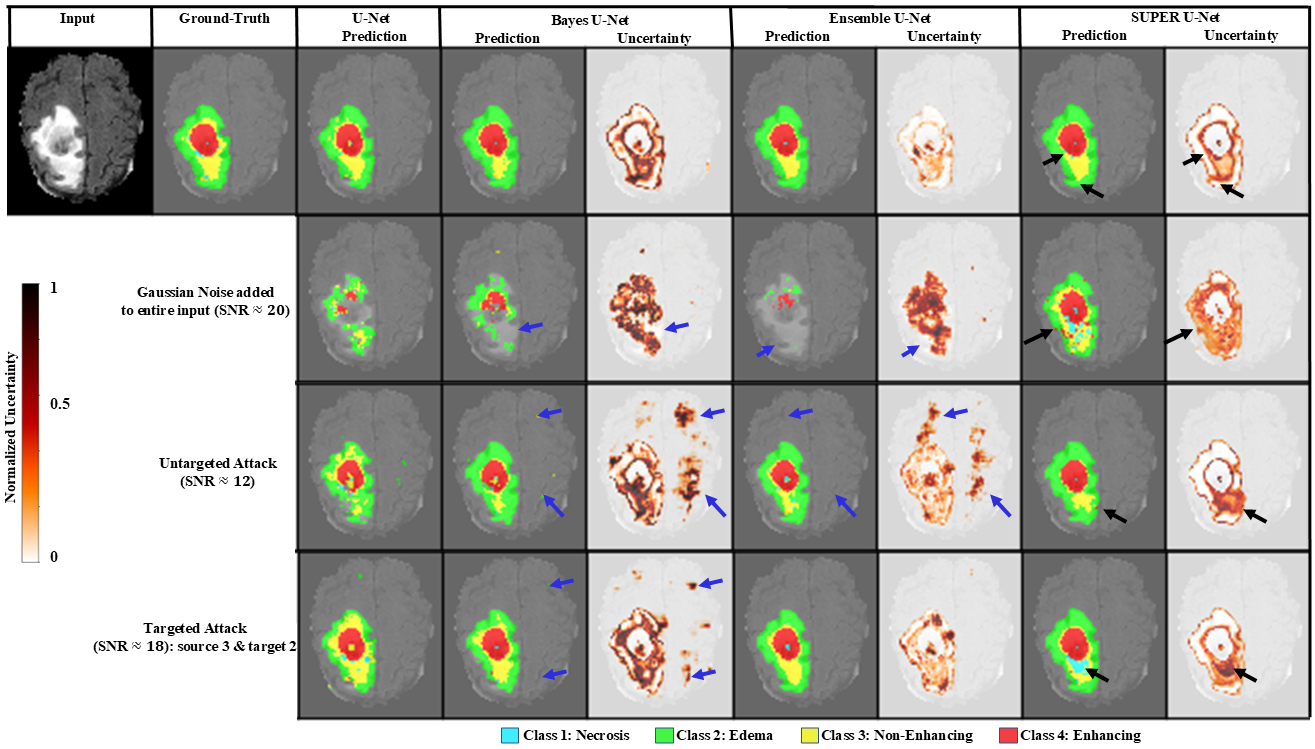}
        \caption{BraTS test data}
        \label{fig:brain}
    \end{subfigure}
    
    \caption{Segmentation of the (a) hippocampus and (b) BraTS test data. The first row shows (left to right) the input image, the ground-truth segmentation, and noise-free models' segmentation predictions (U-Net, Bayes U-Net, Ensemble U-Net, and SUPER U-Net). The uncertainty map of each Bayesian model is shown next to the corresponding prediction. Rows $2, 3, 4$ display the segmented predictions along with their uncertainty maps (when applicable) for additive Gaussian noise and two adversarial attacks, respectively. The black arrows point to regions incorrectly classified by the network. Observe that the corresponding pixels in the uncertainty maps reflect low confidence or higher variance values. The blue arrows refer to inconsistent uncertainty estimates: low confidence is associated with incorrect predictions or high uncertainty for correctly classified regions.}
    \label{fig:combined_hippo_brats}
\end{figure*}

\begin{figure*} 
\begin{center}
\hspace{0cm}\includegraphics[width=.6\textwidth]{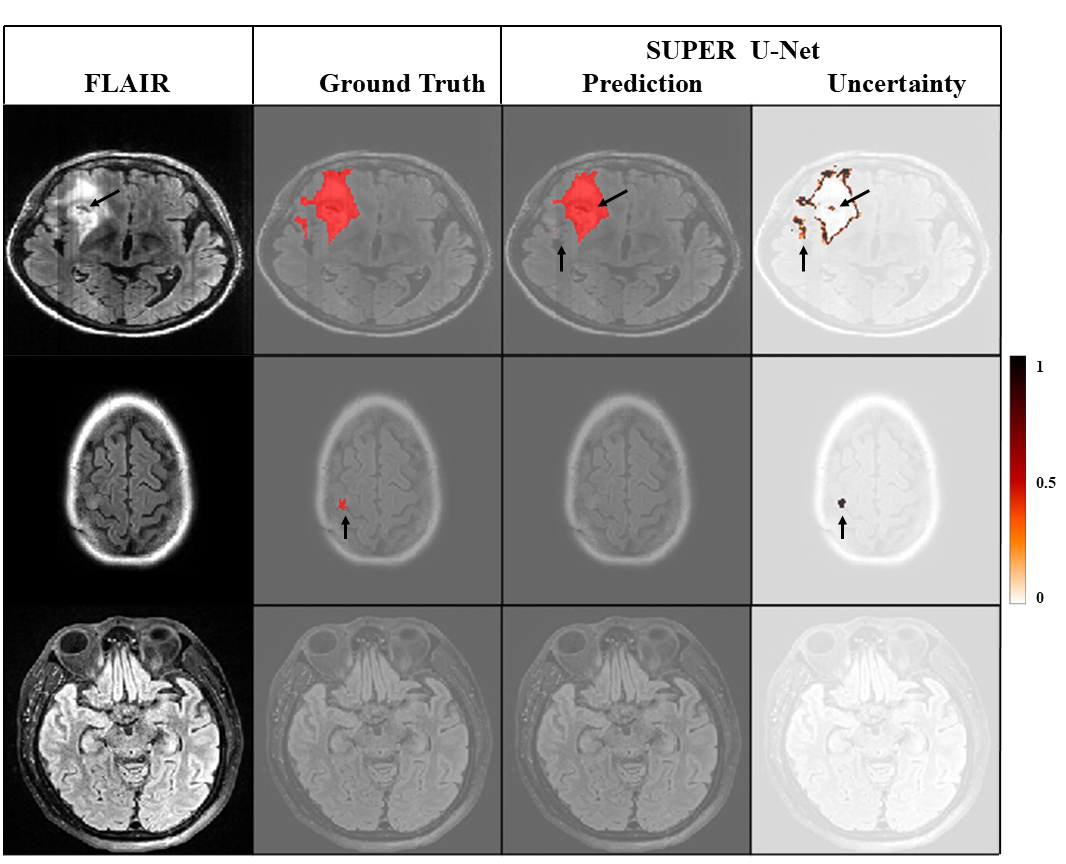}
\caption{Sample scans from the clinical data. Images show (left to right) the Flair input image, the ground-truth segmentation, SUPER U-Net prediction and uncertainty map overlaid on the input scan. The black arrows point to regions incorrectly classified by the network or unusually low (atypical) signals in the FLAIR.}\label{fig:fig_uab}
\end{center}
\end{figure*}

\subsection{Evaluation of the Clinical Data} 
We show that SUPER U-Net can scale to real-world datasets. SUPER U-Net is able to achieve $86 \%$ DSC on held-on test data. Figure \ref{fig:fig_uab} shows sample scans from the UAB clinical data (first column) along with ground-truth segmentation (second column) and SUPER U-Net’s segmentation and associated uncertainty maps (third and fourth columns, respectively). The representative images show that SUPER U-Net is uncertain when a tumor region is missed (scans 1 and 2), as well as for unusually low signal pixels within the tumor (scan 1). A typical tumor is associated with a high FLAIR signal; in the first scan, the central part of the tumor is associated with a low signal, which is atypical (see right arrow in Fig. 6, row 1, column 4). In a sense, the model attracts the physician’s attention to these regions in the image so that they can confirm whether these are part of the tumor or not. In the second scan, the tumor is totally missed, but the model exhibits high uncertainty in the missed region. The last scan has no tumor, and SUPER-Net correctly predicts true negative cases and associates a very low uncertainty (predictive variance $\approx$ 0) or equivalently a high confidence in these predictions.

\subsection{Uncertainty Maps and Predictive Variance -- Quantitative Analysis}
\subsubsection{Uncertainty Maps}
The output of SUPER U-Net consists of the pair: segmentation map (prediction) and uncertainty map (obtained from the predictive covariance). For the other approaches, uncertainty is evaluated through multiple forward passes.
In Fig.\ref{fig:lungs_images}, we present a representative case for the Lungs dataset. In Fig. \ref{fig:combined_hippo_brats} we present representative cases selected from the hippocampus (\ref{fig:hipp}) and BraTS (\ref{fig:brain})  test data. We show the input modality (only FLAIR for the BraTS data), the ground-truth label, and predictions with associated uncertainty maps. The first row presents the noise-free case, the second row reports the predictions and uncertainty maps for the Gaussian noise case, and the third and fourth rows show two examples of adversarial attacks. We normalized the predictive variance of SUPER U-Net for better visual comparison to the uncertainty maps of Bayes U-Net and Ensemble U-Net. We point to regions (pixels) incorrectly classified by our network with the black arrows, and we point to the corresponding locations in the uncertainty maps. It is evident from the figure that SUPER U-Net associates high uncertainty with incorrect predictions and pixels belonging to targeted regions. 

\begin{table*}
\centering
\footnotesize
\caption{DSC for test sample from BraTS Data - performance comparison before and after removal of uncertain pixels.}
\label{tab:quantitative_brats}
\resizebox{\textwidth}{!}{
\begin{tabular}{lccccccccc}
\hline \hline
\multicolumn{1}{l|}{}                                                                         & \multicolumn{3}{c|}{\textit{\textbf{Whole}}}                                                                                & \multicolumn{3}{c|}{\textit{\textbf{Core}}}                                                                                 & \multicolumn{3}{c}{\textit{\textbf{Enhancing}}}                                                                            \\ \hline 
\multicolumn{1}{l|}{}  & \multicolumn{1}{c|}{\textbf{\begin{tabular}[c]{@{}c@{}}Bayes\\ U-Net\end{tabular}}} & \multicolumn{1}{c|}{\textbf{\begin{tabular}[c]{@{}c@{}}Ensemble\\ U-Net\end{tabular}}} & \multicolumn{1}{c|}{\textbf{\begin{tabular}[c]{@{}c@{}}SUPER\\ U-net\end{tabular}}} &  \multicolumn{1}{c|}{\textbf{\begin{tabular}[c]{@{}c@{}}Bayes\\ U-Net\end{tabular}}} & \multicolumn{1}{c|}{\textbf{\begin{tabular}[c]{@{}c@{}}Ensemble\\ U-Net\end{tabular}}} & \multicolumn{1}{c|}{\textbf{\begin{tabular}[c]{@{}c@{}}SUPER\\ U-net\end{tabular}}} &  \multicolumn{1}{c|}{\textbf{\begin{tabular}[c]{@{}c@{}}Bayes\\ U-Net\end{tabular}}} & \multicolumn{1}{c|}{\textbf{\begin{tabular}[c]{@{}c@{}}Ensemble\\ U-Net\end{tabular}}} & \multicolumn{1}{c}{\textbf{\begin{tabular}[c]{@{}c@{}}SUPER\\ U-net\end{tabular}}} \\ 
\hline \hline

\multicolumn{10}{c}{Noise Free }                                                                \\ \hline

\multicolumn{1}{l|}{Original}                                                                           & \multicolumn{1}{c|}{.96}  &   \multicolumn{1}{c|}{.96}                                        &            \multicolumn{1}{c|}{.97}                       & \multicolumn{1}{c|}{.85} 
& \multicolumn{1}{c|}{.87}                                        & \multicolumn{1}{c|}{.89}                      & \multicolumn{1}{c|}{.97}                & \multicolumn{1}{c|}{.96}                                        & \multicolumn{1}{c}{.98}     \\ \hline
\multicolumn{1}{l|}{Uncertain Pixels Removed}                                                                     & \multicolumn{1}{c|}{.99 $\uparrow$}                 & \multicolumn{1}{c|}{.99 $\uparrow$}                         & \multicolumn{1}{c|}{1 $\uparrow$}                                     & \multicolumn{1}{c|}{.91 $\uparrow$}           & \multicolumn{1}{c|}{.88 $\uparrow$}                               & \multicolumn{1}{c|}{.90 $\uparrow$}                     & \multicolumn{1}{c|}{1 $\uparrow$}                & \multicolumn{1}{c|}{.98 $\uparrow$}                                        & \multicolumn{1}{c}{1 $\uparrow$}     \\ \hline

\multicolumn{10}{c}{Gaussian Noise added to entire input (SNR $\approx 20$)}                     \\ \hline

\multicolumn{1}{l|}{Original}                                                                         & \multicolumn{1}{c|}{.56}                & \multicolumn{1}{c|}{.14}    & \multicolumn{1}{c|}{.97}                     & \multicolumn{1}{c|}{.40}                & \multicolumn{1}{c|}{.16}     & \multicolumn{1}{c|}{.86}                  & \multicolumn{1}{c|}{.58}                &  \multicolumn{1}{c|}{.3}    & \multicolumn{1}{c}{.86}     \\ \hline
\multicolumn{1}{l|}{Uncertain Pixels Removed}                                                                        & \multicolumn{1}{c|}{.11$\downarrow$}                &  \multicolumn{1}{c|}{0 $\downarrow$}    & \multicolumn{1}{c|}{.98 $\uparrow$}     & \multicolumn{1}{c|}{.40}                              &  \multicolumn{1}{c|}{.10$\downarrow$}    & \multicolumn{1}{c|}{.99 $\uparrow$}     & \multicolumn{1}{c|}{.61 $\uparrow$}                  & \multicolumn{1}{c|}{.21 $\downarrow$}          &    \multicolumn{1}{c}{1 $\uparrow$}     \\ \hline

\multicolumn{10}{c}{Untargeted adversarial attacks (SNR $\approx 12$) }                                                                                                                             \\ \hline

\multicolumn{1}{l|}{Original}                                                                           & \multicolumn{1}{c|}{.90}                &  \multicolumn{1}{c|}{.91}    & \multicolumn{1}{c|}{.94}                    & \multicolumn{1}{c|}{.74}                &  \multicolumn{1}{c|}{.81}    & \multicolumn{1}{c|}{.78}     & \multicolumn{1}{c|}{.89}                  &          \multicolumn{1}{c|}{.91}        & \multicolumn{1}{c}{.95}     \\ \hline
\multicolumn{1}{l|}{Uncertain Pixels Removed}                                                                           & \multicolumn{1}{c|}{.88 $\downarrow$}                &  \multicolumn{1}{c|}{.79 $\downarrow$}    & \multicolumn{1}{c|}{.95 $\uparrow$}     & \multicolumn{1}{c|}{.75 $\uparrow$}                                &  \multicolumn{1}{c|}{.89 $\uparrow$}    & \multicolumn{1}{c|}{.81 $\uparrow$}     & \multicolumn{1}{c|}{1 $\uparrow$}                  & \multicolumn{1}{c|}{.90 $\downarrow$}          & \multicolumn{1}{c}{1 $\uparrow$}     \\ \hline

\multicolumn{10}{c}{Targeted adversarial attacks (SNR $\approx 18$): source 3, target 2}      \\ \hline

\multicolumn{1}{l|}{Original}                                                                            & \multicolumn{1}{c|}{.93}                &  \multicolumn{1}{c|}{.98}    & \multicolumn{1}{c|}{.99}                    & \multicolumn{1}{c|}{.89}                &  \multicolumn{1}{c|}{.90}    & \multicolumn{1}{c|}{.89}     & \multicolumn{1}{c|}{.95}                  &            \multicolumn{1}{c|}{.96}    &   \multicolumn{1}{c}{.97}     \\ \hline
\multicolumn{1}{l|}{Uncertain Pixels Removed}                                                                           & \multicolumn{1}{c|}{.90 $\downarrow$}                &  \multicolumn{1}{c|}{.98}    & \multicolumn{1}{c|}{.99}     & \multicolumn{1}{c|}{.96 $\uparrow$}                               &  \multicolumn{1}{c|}{.96 $\uparrow$}    & \multicolumn{1}{c|}{.95 $\uparrow$}                    & \multicolumn{1}{c|}{1 $\uparrow$}     &        \multicolumn{1}{c|}{.98 $\uparrow$}      & \multicolumn{1}{c}{1 $\uparrow$}     \\ \hline

\end{tabular}}
\end{table*}

\begin{table} [htb]
\footnotesize
\centering
\caption{SUPER U-Net Predictive Variance for BraTS Dataset.}
\label{tab:var_brats}
\begin{tabular}{l|c|c|c}
 \hline \hline
\multicolumn{1}{l|}{}                     & \multicolumn{1}{c|}{\textbf{Whole}} &\multicolumn{1}{c|}{\textbf{Core}} &  \multicolumn{1}{c}{\textbf{Enhancing}} \\ \hline \hline

\multicolumn{1}{l|}{Correct}     &.007  &.008  &.008      \\ \hline

\multicolumn{1}{l|}{Incorrect} &.289  &.307  & \multicolumn{1}{c}{.336}            \\ \hline

\end{tabular}
\end{table}
\begin{figure*} [t]
\begin{center}
\includegraphics[width=1.\textwidth]{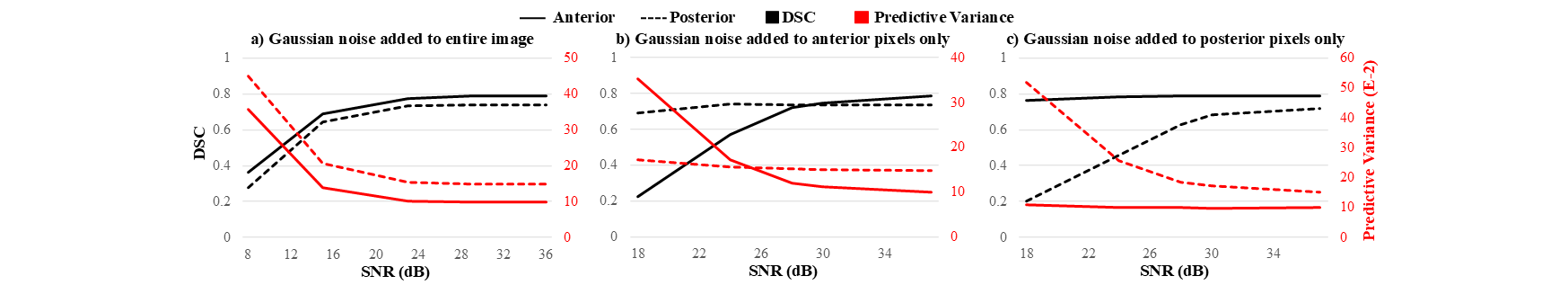}
\caption{Accuracy, measured by Dice Similarity Coefficient (DSC) and plotted in black, and average predictive variance of SUPER U-Net, plotted in red, under various levels of Gaussian noise added in the test data for hippocampus dataset. SNR denotes the signal-to-noise ratio. (a) Noise is added to the entire input. (b) Noise is added to the anterior pixels only. (c) Noise is added to the posterior pixels only.} \label{fig:var_dsc_gaussian_hip}
\end{center}
\end{figure*}
\subsubsection{Predictive Variance -- Quantitative Analysis} 
\begin{figure*}
\begin{center}
\includegraphics[width=1\textwidth]{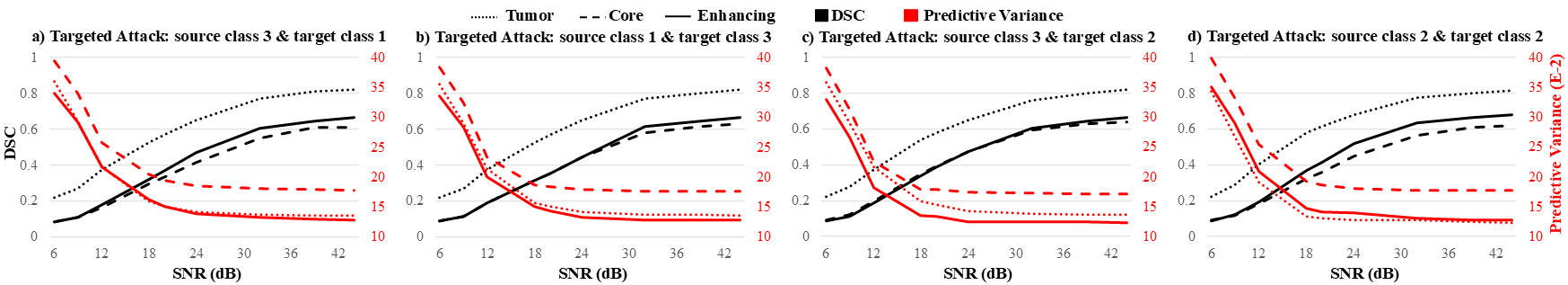}
\caption{Accuracy, measured by Dice Similarity Coefficient (DSCs) and plotted in black, and average predictive variance of SUPER U-Net, plotted in red, under various levels of adversarial attacks applied to the test data for BraTS dataset.  SNR denotes the signal-to-noise ratio. Test data is corrupted with targeted attacks: (a) source class 3 and target class 1, (b) source class 1 and target class 3, (c) source class 3 and target class 2, (d) source class 2 and target class 3.}
\label{fig:adv_DSC_var}
\end{center}
\end{figure*}

We investigate the response of the derived second moment (variance/uncertainty) and relate it to the model’s performance (DSC). 
We calculate the average predictive variance from uncertainty maps and plot these values against various levels of Gaussian noise in Fig. \ref{fig:var_dsc_gaussian_hip}, and adversarial attacks in Fig. \ref{fig:adv_DSC_var}, for hippocampus and BraTS datasets. It is more instructive and insightful if sub-plots in both figures are interpreted from right to left, i.e., decreasing SNR or equivalently increasing noise in the test data. We note that the predictive variance increases monotonically with increasing noise (i.e., decreasing SNR) for all three sub-figures in Fig. \ref{fig:var_dsc_gaussian_hip} and all four sub-figures in Fig. \ref{fig:adv_DSC_var}. This behavior, i.e., increasing uncertainty with increasing noise, demonstrates that the network is aware of higher noise in the input. A useful and meaningful uncertainty estimate should convey a lower confidence/ higher uncertainty for low-accuracy segmented images \citep{mehta2020uncertainty}. Table \ref{tab:var_brats} reports SUPER U-Net average predictive variance for correctly classified and misclassified pixels on noise-free BraTS test set. Observe that the incorrect pixels are associated to high variance or less confident predictions.

Following the quantitative uncertainty evaluation task in the BRATS challenge \citep{mehta2020uncertainty}, we compute the percentage change in DSCs when \emph{uncertain} pixels are removed, and DSC is computed only using the remaining pixels. To define \emph{uncertain} pixels, for each model, we set as a threshold the average predictive uncertainty for correctly classified pixels for the noise-free case. All pixels with an uncertainty value above this threshold are marked as uncertain and removed from the computation of the DSC. We report the change in DSCs in table \ref{tab:quantitative_brats}. The sample scan corresponds to that provided qualitatively in Fig. \ref{fig:brain}.  Our approach consistently produces higher (↑) DSCs after removing uncertain pixels, i.e., unlike other approaches, our predictive variance (uncertainty) is above the threshold only for incorrectly classified pixels. Such information is valuable for detecting when the network may fail and its predictions may become untrustworthy.

\section{Discussion}\label{discussion}
\subsection{The Significance of Uncertainty Information}
The reliability of segmentation predictions can be assessed by examining the associated uncertainty maps. For SUPER U-Net, higher uncertainty values are concentrated in regions where the segmentation is incorrect or where the input is perturbed by noise or adversarial attacks (Figures~\ref{fig:hipp}, \ref{fig:brain}, black arrows). When the predicted segmentation is accurate, uncertainty is largely confined to object boundaries. In contrast, Bayes U-Net and Ensemble U-Net often produce inconsistent estimates (Figures~\ref{fig:hipp}, \ref{fig:brain}, blue arrows), sometimes assigning low uncertainty to incorrect predictions and/or high uncertainty to correctly segmented regions. These results demonstrate that explicitly propagating covariance information during training leads to uncertainty estimates that better reflect segmentation reliability.

\subsection{Clinical Impact}
Precise and consistent segmentation remains a critical requirement in clinical workflows.
DL segmentation methods offer promise by providing objective delineations of tumors and surrounding anatomy. However, their lack of reliability and consistency has limited adoption in the clinic  \citep{carannante2023bayesian}. For instance, models may encounter tumor morphologies not represented in training data, image artifacts from different acquisition protocols, or even adversarial perturbations \citep{finlayson2019adversarial}. In such cases, model uncertainty estimates can play a key role in mitigating risks by alerting clinicians to regions where predictions are less reliable.

The proposed SUPER U-Net produces both a segmentation map and a corresponding pixel-level uncertainty map, enabling users to identify regions where predictions may be less reliable. Rather than treating model outputs as definitive, this approach supports more informed human–AI interaction by directing attention to areas of low confidence. Uncertainty estimates may also support downstream applications such as threshold-based triaging, where cases with high uncertainty are flagged for additional human review. While this has the potential to enhance trust in DL-assisted tumor monitoring and treatment planning, further prospective validation studies will be needed before deployment in routine clinical settings.

\begin{table}[t]
\caption{Inference time per image}
\footnotesize
\label{tab:time}
\centering
\begin{tabular}{l|p{2.2cm}|p{2.2cm}|p{2.2cm}|p{2.2cm}}
\hline \hline
  & \textbf{U-Net} & \textbf{Bayes U-Net} & \textbf{Ensemble U-Net} & \textbf{SUPER U-Nets} \\ \hline \hline
Time (min) & 0.81             & 0.82 $N_{1}^*$   & 0.81 $N_{2}^*$   & 1.92 \\ \hline
\end{tabular}
\begin{flushleft}
    $^*N_{1}$ and $N_{2}$ denote the number of runs at inference time and ensemble networks, respectively.
\end{flushleft}
\end{table}

\subsection{Computational Considerations and Limitations}
\subsubsection{Computational Complexity}  
We report the average inference time of all four networks in Table \ref{tab:time}. SUPER U-Net requires almost twice the time to process a single image at inference compared to a deterministic U-Net. This increase is due to the propagation of the covariance information, which involves additional operations. 

Other approaches that deliver uncertainty, i.e.,  Bayes U-Net and Ensemble U-Net, take the same time as that of a deterministic U-Net for one pass (or one model). However, these approaches necessitate multiple passes to calculate the variance of the prediction. For example, Bayes U-Net with $N=20$   rewuires $16.4$ ms for each image, nearly 8 times more than SUPER U-Net. Ensemble methods also demand extended training time and storage to save multiple models. 

\subsubsection{Storage}  
The computational complexity of the proposed SUPER U-Net framework is primarily driven by operations on the mean and variance vectors, not by parameter storage. Given the tensor normal distribution, SUPER U-Net requires one additional trainable parameter per convolutional kernel \citep{dera2019extended}, leaving the overall parameter count close to that of a deterministic model.  

For instance, on the Lungs dataset with kernel size $3 \times 3$, the deterministic model has $n_1 = 1440$ parameters ($\sim 5.625$ KB), while SUPER U-Net has $n_2 = 1600$ parameters ($\sim 6.25$ KB). This modest increase in storage highlights that the main computational burden arises from performing separate operations on the mean and variance vectors rather than from storing the additional parameters.

\subsubsection{Limitations}  
Despite these advantages, several limitations must be acknowledged. One key challenge concerns its performance on small and imbalanced datasets. While Bayesian methods, including early works such as \citep{blundell2015weight} and \citep{kendall2017bayesian}, have demonstrated improved generalization in low-data regimes, dataset imbalance may influence uncertainty quality, especially when rare structures are underrepresented during training. 

Second, SUPER U-Net is sensitive to prior assumptions and hyperparameter choices. The Gaussian prior over network weights, while common, may not optimally capture complex medical imaging distributions. Likewise, hyperparameters such as the prior variance and the KL regularization term  significantly influence model behavior, and suboptimal tuning could lead to failure in learning, overconfident predictions, or excessive uncertainty. Future work could explore alternative priors, propagation of higher-order moments, and targeted evaluations on rare disease segmentation and other data-scarce applications.

\subsection{Strengths and Contribution}

\subsubsection{Trade-offs and Practical Advantages}  
Although SUPER U-Net incurs higher inference time than a deterministic U-Net, it remains substantially more efficient than Bayes or Ensemble U-Net while delivering superior robustness to noise and adversarial attacks. The model’s intrinsic uncertainty estimates provide an effective way to assess prediction reliability. Prior work has also shown that Bayesian models can identify redundant kernels that may be pruned without accuracy loss, potentially reducing storage demands \citep{carannante2020self}.  

\subsubsection{Key Strengths}
SUPER U-Net offers several advantages over commonly used state-of-the-art uncertainty quantification approaches such as Bayes U-Net and Ensemble U-Net:
\begin{itemize}
    \item Intrinsic uncertainty learning: Mean and covariance are propagated during training, in contrast to post-hoc uncertainty estimation or sampling-based approaches.
    \item Increased Robustness: Demonstrates superior performance under Gaussian noise and adversarial attacks, especially at high levels of noise and complex tasks/data such as BraTS segmentation.
    \item Generalizability: Performs consistently across multiple datasets and noise conditions, highlighting its broad applicability.
    
    \item Computational efficiency: Avoids repeated forward passes or large ensembles, reducing inference cost while improving predictive reliability.
\end{itemize}
\section{Conclusion}\label{conclude}
This study introduced SUPER-Net, a novel Bayesian DL framework that effectively quantifies uncertainty in medical image segmentation tasks using encoder-decoder architectures. 
The key contributions of SUPER-Net are:  
\begin{itemize}
    \item The ability to produce pixel-wise uncertainty maps alongside segmentation outputs in real-time without relying on expensive post-hoc sampling techniques like Monte Carlo.  
    \item Robust performance across diverse datasets and noise conditions, with resilience to adversarial attacks.  
    \item Meaningful uncertainty estimates that provide actionable guidance for trust and decision-making in medical imaging.  
\end{itemize}

Researchers and practitioners in medical imaging can benefit from SUPER-Net by utilizing its uncertainty maps to improve the reliability of automated segmentation, especially in high-risk applications. This framework can be adapted for different architectures and imaging modalities, thus serving as a valuable tool for applications requiring high levels of confidence in predictions.

While SUPER-Net achieves strong performance, it also introduces trade-offs. First, propagating both means and covariances through the network increases computational cost relative to deterministic models, which may limit scalability in resource-constrained environments. Moreover, its performance depends on prior assumptions and hyperparameter choices: SUPER-Net currently adopts a Gaussian prior over network weights, a standard but potentially restrictive assumption for complex medical imaging tasks. These aspects represent both limitations and opportunities for refinement.  

For future work, we plan to (i) explore more expressive prior and posterior distributions beyond the Gaussian assumption to better capture uncertainties in heterogeneous data, (ii) investigate strategies for hyperparameter optimization to reduce sensitivity to manual tuning, and (iii) develop methods to mitigate computational overhead to improve scalability and facilitate clinical integration.


\begin{thebibliography}{10}
\expandafter\ifx\csname url\endcsname\relax
  \def\url#1{\texttt{#1}}\fi
\expandafter\ifx\csname urlprefix\endcsname\relax\def\urlprefix{URL }\fi
\expandafter\ifx\csname href\endcsname\relax
  \def\href#1#2{#2} \def\path#1{#1}\fi

\bibitem{liu2019comparison}
X.~Liu, L.~Faes, A.~U. Kale, S.~K. Wagner, D.~J. Fu, A.~Bruynseels, T.~Mahendiran, G.~Moraes, M.~Shamdas, C.~Kern, et~al., A comparison of deep learning performance against health-care professionals in detecting diseases from medical imaging: a systematic review and meta-analysis, The lancet digital health 1~(6) (2019) e271--e297.

\bibitem{biggio2018wild}
B.~Biggio, F.~Roli, Wild patterns: Ten years after the rise of adversarial machine learning, Pattern Recognition 84 (2018) 317--331.

\bibitem{goodfellow2015explaining}
I.~J. Goodfellow, J.~Shlens, C.~Szegedy, Explaining and harnessing adversarial examples, in: International Conference on Learning Representations (ICLR), 2015.

\bibitem{finlayson2019adversarial}
S.~G. Finlayson, J.~D. Bowers, J.~Ito, J.~L. Zittrain, A.~L. Beam, I.~S. Kohane, Adversarial attacks on medical machine learning, Science 363~(6433) (2019) 1287--1289.

\bibitem{goan2020bayesian}
E.~Goan, C.~Fookes, Bayesian neural networks: An introduction and survey, in: Case Studies in Applied Bayesian Data Science, Springer, 2020, pp. 45--87.

\bibitem{dera2019extended}
D.~Dera, G.~Rasool, N.~Bouaynaya, Extended {V}ariational {I}nference for {P}ropagating {U}ncertainty in {C}onvolutional {N}eural {N}etworks, in: 2019 IEEE 29th International Workshop on Machine Learning for Signal Processing (MLSP), IEEE, 2019, pp. 1--6.

\bibitem{doucet2009tutorial}
A.~Doucet, A.~M. Johansen, A tutorial on particle filtering and smoothing: Fifteen years later, Handbook of Nonlinear Filtering 12 (2009) 656--704.

\bibitem{long2015fully}
J.~Long, E.~Shelhamer, T.~Darrell, Fully convolutional networks for semantic segmentation, in: Proceedings of the IEEE conference on Computer Vision and Pattern Recognition, 2015, pp. 3431--3440.

\bibitem{ronneberger2015u}
O.~Ronneberger, P.~Fischer, T.~Brox, U-net: Convolutional networks for biomedical image segmentation, in: International Conference on Medical image computing and computer-assisted intervention, Springer, 2015, pp. 234--241.

\bibitem{cao2024rasnet}
G.~Cao, Z.~Sun, C.~Wang, H.~Geng, H.~Fu, Z.~Yin, M.~Pan, Rasnet: Renal automatic segmentation using an improved u-net with multi-scale perception and attention unit, Pattern Recognition (2024) 110336.

\bibitem{cahall2019inception}
D.~E. Cahall, G.~Rasool, N.~C. Bouaynaya, H.~M. Fathallah-Shaykh, Inception modules enhance brain tumor segmentation, Frontiers in computational neuroscience 13 (2019) 44.

\bibitem{rehman2020bu}
M.~U. Rehman, S.~Cho, J.~H. Kim, K.~T. Chong, Bu-net: Brain tumor segmentation using modified u-net architecture, Electronics 9~(12) (2020) 2203.

\bibitem{yan20233d}
Q.~Yan, S.~Liu, S.~Xu, C.~Dong, Z.~Li, J.~Q. Shi, Y.~Zhang, D.~Dai, 3d medical image segmentation using parallel transformers, Pattern Recognition 138 (2023) 109432.

\bibitem{thisanke2023semantic}
H.~Thisanke, C.~Deshan, K.~Chamith, S.~Seneviratne, R.~Vidanaarachchi, D.~Herath, Semantic segmentation using vision transformers: A survey, Engineering Applications of Artificial Intelligence 126 (2023) 106669.

\bibitem{xiao2023transformers}
H.~Xiao, L.~Li, Q.~Liu, X.~Zhu, Q.~Zhang, Transformers in medical image segmentation: A review, Biomedical Signal Processing and Control 84 (2023) 104791.

\bibitem{kirillov2023segment}
A.~Kirillov, E.~Mintun, N.~Ravi, H.~Mao, C.~Rolland, L.~Gustafson, T.~Xiao, S.~Whitehead, A.~C. Berg, W.-Y. Lo, et~al., Segment anything, in: Proceedings of the IEEE/CVF International Conference on Computer Vision, 2023, pp. 4015--4026.

\bibitem{huang2024segment}
Y.~Huang, X.~Yang, L.~Liu, H.~Zhou, A.~Chang, X.~Zhou, R.~Chen, J.~Yu, J.~Chen, C.~Chen, et~al., Segment anything model for medical images?, Medical Image Analysis 92 (2024) 103061.

\bibitem{gal2016dropout}
Y.~Gal, Z.~Ghahramani, Dropout as a {B}ayesian {A}pproximation: {R}epresenting {M}odel {U}ncertainty in {D}eep {L}earning, in: International Conference on Machine Learning, 2016, pp. 1050--1059.

\bibitem{lakshminarayanan2017simple}
B.~Lakshminarayanan, A.~Pritzel, C.~Blundell, {S}imple and {S}calable {P}redictive {U}ncertainty {E}stimation using {D}eep {E}nsembles, in: Advances in Neural Information Processing Systems, 2017, pp. 6402--6413.

\bibitem{kendall2017bayesian}
A.~Kendall, V.~Badrinarayanan, R.~Cipolla, Bayesian segnet: Model uncertainty in deep convolutional encoder-decoder architectures for scene understanding, in: Proceedings of the British Machine Vision Conference (BMVC), 2017.

\bibitem{jungo2017towards}
A.~Jungo, R.~McKinley, R.~Meier, U.~Knecht, L.~Vera, J.~P{\'e}rez-Beteta, D.~Molina-Garc{\'\i}a, V.~M. P{\'e}rez-Garc{\'\i}a, R.~Wiest, M.~Reyes, Towards {U}ncertainty-{A}ssisted {B}rain {T}umor {S}egmentation and {S}urvival {P}rediction, in: International MICCAI Brainlesion Workshop, Springer, 2017, pp. 474--485.

\bibitem{roy2019bayesian}
A.~G. Roy, S.~Conjeti, N.~Navab, C.~Wachinger, A.~D.~N. Initiative, et~al., Bayesian quicknat: Model uncertainty in deep whole-brain segmentation for structure-wise quality control, NeuroImage 195 (2019) 11--22.

\bibitem{larrazabal2021orthogonal}
A.~J. Larrazabal, C.~Mart{\'\i}nez, J.~Dolz, E.~Ferrante, Orthogonal ensemble networks for biomedical image segmentation, in: International Conference on Medical Image Computing and Computer-Assisted Intervention, Springer, 2021, pp. 594--603.

\bibitem{kamnitsas2017ensembles}
K.~Kamnitsas, W.~Bai, E.~Ferrante, S.~McDonagh, M.~Sinclair, N.~Pawlowski, M.~Rajchl, M.~Lee, B.~Kainz, D.~Rueckert, et~al., Ensembles of multiple models and architectures for robust brain tumour segmentation, in: International MICCAI brainlesion workshop, Springer, 2017, pp. 450--462.

\bibitem{ghoshal2021estimating}
B.~Ghoshal, A.~Tucker, B.~Sanghera, W.~Lup~Wong, Estimating uncertainty in deep learning for reporting confidence to clinicians in medical image segmentation and diseases detection, Computational Intelligence 37~(2) (2021) 701--734.

\bibitem{baumgartner2019phiseg}
C.~F. Baumgartner, K.~C. Tezcan, K.~Chaitanya, A.~M. H{\"o}tker, U.~J. Muehlematter, K.~Schawkat, A.~S. Becker, O.~Donati, E.~Konukoglu, Phiseg: Capturing uncertainty in medical image segmentation, in: International Conference on Medical Image Computing and Computer-Assisted Intervention, Springer, 2019, pp. 119--127.

\bibitem{li2023region}
H.~Li, Y.~Nan, J.~Del~Ser, G.~Yang, Region-based evidential deep learning to quantify uncertainty and improve robustness of brain tumor segmentation, Neural Computing and Applications 35~(30) (2023) 22071--22085.

\bibitem{guo2024uctnet}
X.~Guo, X.~Lin, X.~Yang, L.~Yu, K.-T. Cheng, Z.~Yan, Uctnet: Uncertainty-guided cnn-transformer hybrid networks for medical image segmentation, Pattern Recognition 152 (2024) 110491.

\bibitem{graves2011practical}
A.~Graves, {P}ractical {V}ariational {I}nference for {N}eural {N}etworks, in: Advances in neural information processing systems, 2011, pp. 2348--2356.

\bibitem{blundell2015weight}
C.~Blundell, J.~Cornebise, K.~Kavukcuoglu, D.~Wierstra, Weight uncertainty in neural networks, in: Proceedings of the 32nd International Conference on Machine Learning (ICML), 2015, pp. 1613--1622.

\bibitem{ma_jun_2020_3757476}
M.~Jun, G.~Cheng, W.~Yixin, A.~Xingle, G.~Jiantao, Y.~Ziqi, Z.~Minqing, L.~Xin, D.~Xueyuan, C.~Shucheng, W.~Hao, M.~Sen, Y.~Xiaoyu, N.~Ziwei, L.~Chen, T.~Lu, Z.~Yuntao, Z.~Qiongjie, D.~Guoqiang, H.~Jian, {COVID-19 CT Lung and Infection Segmentation Dataset} (Apr. 2020).
\newblock \href {https://doi.org/10.5281/zenodo.3757476} {\path{doi:10.5281/zenodo.3757476}}.

\bibitem{antonelli2022medical}
M.~Antonelli, A.~Reinke, S.~Bakas, K.~Farahani, A.~Kopp-Schneider, B.~A. Landman, G.~Litjens, B.~Menze, O.~Ronneberger, R.~M. Summers, et~al., The medical segmentation decathlon, Nature Communications 13~(1) (2022) 4128.

\bibitem{menze2014multimodal}
B.~H. Menze, A.~Jakab, S.~Bauer, J.~Kalpathy-Cramer, K.~Farahani, J.~Kirby, Y.~Burren, N.~Porz, J.~Slotboom, R.~Wiest, et~al., The multimodal brain tumor image segmentation benchmark (brats), IEEE transactions on medical imaging 34~(10) (2014) 1993--2024.

\bibitem{pmid31136584}
H.~M. Fathallah-Shaykh, A.~DeAtkine, E.~Coffee, E.~Khayat, A.~K. Bag, X.~Han, P.~P. Warren, M.~Bredel, J.~Fiveash, J.~Markert, N.~Bouaynaya, L.~B. Nabors, {{D}iagnosing growth in low-grade gliomas with and without longitudinal volume measurements: {A} retrospective observational study}, PLoS Med 16~(5) (2019) e1002810.

\bibitem{liu2017delving}
Y.~Liu, X.~Chen, C.~Liu, D.~Song, Delving into transferable adversarial examples and black-box attacks, in: International Conference on Learning Representations (ICLR), 2017.

\bibitem{madry2018towards}
A.~Madry, A.~Makelov, L.~Schmidt, D.~Tsipras, A.~Vladu, Towards deep learning models resistant to adversarial attacks, in: International Conference on Learning Representations (ICLR), 2018.

\bibitem{mehta2020uncertainty}
R.~Mehta, A.~Filos, Y.~Gal, T.~Arbel, Uncertainty evaluation metric for brain tumour segmentation, in: Medical Imaging with Deep Learning (MIDL), 2020.

\bibitem{carannante2023bayesian}
G.~Carannante, N.~C. Bouaynaya, Bayesian deep learning detection of anomalies and failure: Application to medical images, in: 2023 IEEE 33rd International Workshop on Machine Learning for Signal Processing (MLSP), IEEE, 2023, pp. 1--6.

\bibitem{carannante2020self}
G.~Carannante, D.~Dera, G.~Rasool, N.~C. Bouaynaya, Self-compression in bayesian neural networks, in: 2020 IEEE 30th International Workshop on Machine Learning for Signal Processing (MLSP), IEEE, 2020, pp. 1--6.

\end{thebibliography}

\end{document}